# What are the limits to biomedical research acceleration through general-purpose AI?

Konstantin Hebenstreit[1]†, Constantin Convalexius[1]†, Stephan Reichl[1,2]†, Stefan Huber[1], Christoph Bock[1,2], Matthias Samwald[1,3]*

[1] Medical University of Vienna, Institute of Artificial Intelligence, Center for Medical Data Science, Vienna, Austria
[2] CeMM Research Center for Molecular Medicine of the Austrian Academy of Sciences, Vienna, Austria
[3] *Accelerate Europe* Initiative
† These authors contributed equally to this work.
* Corresponding author: matthias.samwald [at] meduniwien.ac.at

## Abstract

Although general-purpose artificial intelligence (GPAI) is widely expected to accelerate scientific discovery, its practical limits in biomedicine remain unclear. We assess this potential by developing a framework of GPAI capabilities across the biomedical research lifecycle. Our scoping literature review indicates that current GPAI could deliver a speed increase of around 2x, whereas future GPAI could facilitate strong acceleration of up to 25x for physical tasks and 100x for cognitive tasks. However, achieving these gains may be severely limited by factors such as irreducible biological constraints, research infrastructure, data access, and the need for human oversight. Our expert elicitation with eight senior biomedical researchers revealed skepticism regarding the strong acceleration of tasks such as experiment design and execution. In contrast, strong acceleration of manuscript preparation, review and publication processes was deemed plausible. Notably, all experts identified the assimilation of new tools by the scientific community as a critical bottleneck. Realising the potential of GPAI will therefore require more than technological progress; it demands targeted investment in shared automation infrastructure and systemic reforms to research and publication practices.

## Introduction

Artificial intelligence affects large parts of human society, yet its most profound impact may lie in **accelerating scientific discovery**. The OECD recently emphasized that enhancing research productivity through AI could be "the most economically and socially valuable" application of this technology[1], echoing Nobel laureate Robert Solow's foundational insight that technological advancement—not merely capital or labor—drives sustainable economic prosperity[2,3]. Influential voices such as Nobel laureate Demis Hassabis and the World Economic Forum similarly underscore that the primary value of AI lies in accelerating science[4].

AI models trained on specific tasks ('narrow AI') have long been used as part of biomedical research, including breakthroughs such as AlphaFold's[5] protein structure prediction capabilities. The advent of **general-purpose AI (GPAI),** such as large language models (LLMs), marked a significant progression from narrow AI toward more generalized capabilities. GPAI models have been defined as exhibiting "significant generality and are capable of competently performing a wide range of distinct tasks"[6].

Researchers face overwhelming numbers of scientific publications, massive data sets, and the increasing demands of multidisciplinary collaboration[7–11]. GPAI models **integrate knowledge across disciplines**, such as biology, chemistry, engineering, and computing[12]. They demonstrate proficiency **across diverse tasks,** including coding, mathematics, and logical reasoning—achieving or surpassing human-level performance on rigorous scientific and coding benchmarks[13,14]. These capabilities are enhanced by rapid progress in mixture-of-expert architectures[15] and reinforcement learning[16] techniques, leading to sophisticated **reasoning models**[17–19]. Finally, GPAI models can utilize **external tools**—such as search systems, specialized narrow AI systems, or lab automation frameworks[20–22].

GPAI models also drive emerging **autonomous AI agents**, which independently plan, reason, utilize tools, and iteratively explore scientific problems, potentially without significant human oversight. Empirical evidence supports the transformative potential of agent-driven research acceleration[21,23–25]. AI agents quickly distill literature, recognize contradictions and develop new hypotheses[21,26,27]. Combined with continuously operating **self-driving laboratories**, research timelines for projects may shrink from years to months or weeks[28,29]. One notable example is GPAI-driven drug discovery for pulmonary fibrosis, where one company reported reducing the time from discovery to preclinical candidate from at least 5–6 years[30] to 18 months[31]. This >3× acceleration could accelerate the breaking of "Eroom's Law"[32], the trend of exponentially slower and more expensive drug development. However, biomedical research depends on steps that cannot be compressed, setting hard limitations to research acceleration in the life sciences.

Our study estimates the potential acceleration of established biomedical research processes using GPAI within the current research paradigm. Rather than speculating about systemic changes such as replacing biological experiments with in silico alternatives—changes that are inherently difficult to predict and quantify—we focus on how GPAI can automate and accelerate research methods currently employed. This allows us to provide evidence-based, practical estimates of realistic acceleration factors that are relevant for the immediate future, while accounting for systemic or fundamental limitations inherent to biomedical research.

Within this scope, we explore the boundaries of biomedical research acceleration through GPAI, addressing three central questions:

1. What **acceleration factors** may current and future GPAI plausibly achieve across different biomedical research tasks?

2. What challenges and constraints might **limit** this acceleration?

3. How might GPAI **transform research workflows** and what are **key policy considerations** to ensure responsible innovation?

## Results

### Frameworks to assess the acceleration of biomedical research using GPAI

*GPAI capability framework*

To analyze the potential for GPAI-driven acceleration of biomedical research, we first developed a framework of GPAI capability levels. We identified and analyzed four **relevant existing frameworks**: DeepMind's "Levels of AGI", which explains AI system generality by comparing narrow and broad AI performance to human skill levels[33]; SAE International's "Driving automation systems", which describes the progression of human-autonomous vehicle interaction[34]; "AI agents for biomedical discovery", which details increasing levels of AI agency in research[35]; and "Self-Driving Laboratories", which maps the integration of software and hardware processes for increasing laboratory autonomy[36].

Drawing on these frameworks and analyzing biomedical research requirements, we synthesized a **simple, unified framework of GPAI research capability** with two key dimensions: **cognitive capability** and **physical capability**. Cognitive capability encompasses research activities primarily involving information processing, analysis, and decision-making, including literature review, hypothesis generation, experimental design, data analysis, result interpretation, and manuscript preparation. Physical capability involves laboratory procedures, experimental setup, and material handling through robotics, lab automation, and experiment setup through experiment execution.

Concomitant increases in both cognitive and physical capabilities lead to increases in the emergent attribute of *autonomy*, i.e., the ability to operate independently across the complete research cycle. High levels of autonomy are key to the most radical acceleration of research (Figure 1).

For both cognitive and physical capabilities, we define **three levels** that chart the progression of GPAI integration into the research process:

- At the **"No GPAI"** level, humans perform all work manually, possibly assisted by non-GPAI tools.

- **"Next-level" capabilities** represent GPAI systems that partially automate the research cycle but require significant human intervention, as demonstrated by several current systems.

- **"Maximum-level" capabilities** represent a radically transformed future scenario with advanced capabilities and high-level autonomy where GPAI conducts expert-level research with minimal to no human supervision.

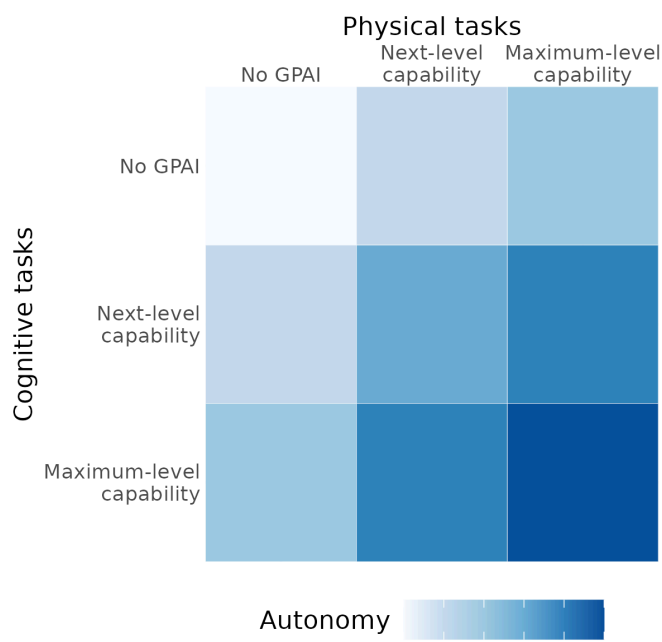

*Figure 1: Capability framework illustrating how cognitive and physical capabilities of GPAI combine to yield autonomy.* Higher autonomy levels enable increasingly significant acceleration of the biomedical research process. Adapted from Tom et al.[36]

*Research task framework*

Drawing on established concepts from the literature[37–41], we developed a structured, end-to-end framework of the biomedical research process comprising nine major research tasks (Figure 2), as well as constituent sub-tasks (detailed in Table S1). This allows us to map specific GPAI capabilities found in the literature to individual research tasks, track the varying levels of automation possible across different aspects of research, and better identify bottlenecks and opportunities for acceleration.

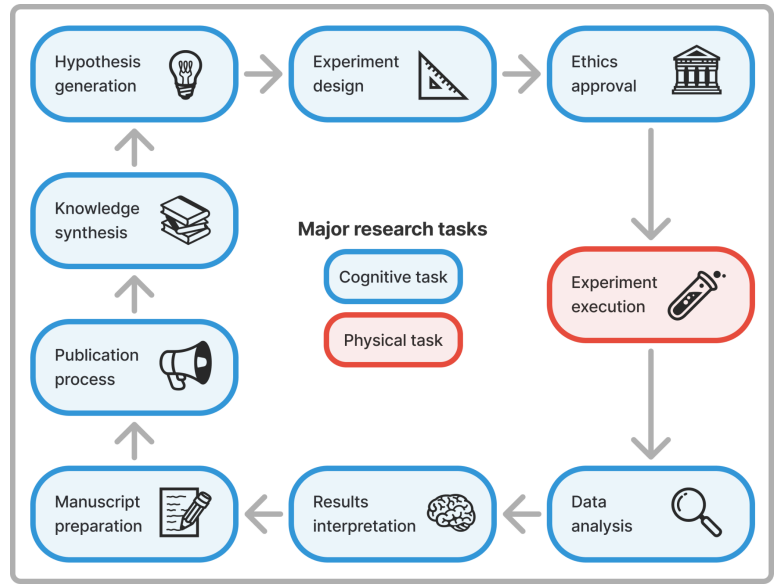

*Figure 2: End-to-end biomedical research task framework covering nine major tasks.* *Blue boxes indicate cognitive tasks, the red box the sole physical task (experiment execution), and arrows depict the typical order and iterative nature of research projects.*

The major research tasks in our framework consist of the following:

**1. Knowledge synthesis** encompasses the collection, critical evaluation and integration of scientific information. Recent GPAI approaches achieve human-level or higher precision in literature-based tasks and operate at high throughput[27]. Systems automate literature search, evaluation, and summarization[22,42] while demonstrating critical thinking and evaluation capabilities[25,43].

**2. Idea & hypothesis generation** involves identifying promising research problems, formulating testable hypotheses, developing theoretical frameworks, and assessing feasibility. GPAI systems are increasingly able to develop new research ideas by analyzing patterns in existing literature and proposing candidate hypotheses and mechanisms[26,31,44]. Advanced systems can autonomously generate and, in some cases, rank candidate hypotheses by novelty and feasibility, then refine key research questions into actionable, experimentally testable goals[24,25,42,45].

**3. Experiment design** includes method selection, protocol development, and planning data collection, including quality control. GPAI systems now demonstrate capabilities in suggesting experimental methods and designing wet-lab or computational protocols[25,26,46], with advanced systems able to formulate experimental plans, predict and optimize parameters, and integrate quality control and reproducibility measures, including protocol standardization[22,42,47–49].

**4. Ethics approval & permits** involve initial screening of research proposals, scientific review, ethics assessment, regulatory compliance, and administrative processing. While currently the most human-centered process, GPAI is beginning to assist in areas such as automated screening for ethical issues, documentation preparation, and compliance checks[50–52]. GPAI systems have potential to enhance informed consent processes[53,54], improve scientific review efficiency, and support compliance checks against applicable regulations and guidelines[55].

**5. Experiment execution** represents the physical dimension of research, including preparation, execution, documentation, troubleshooting, material management, and equipment maintenance. Self-driving labs can execute fully automated multi-step workflows with parallel samples, achieving significant acceleration through continuous operation, exception handling and adaptive optimization[28,48,56–58]. Advanced robotic systems execute protocols, monitor experimental conditions, and manage samples while improving throughput and reducing time requirements[29,48,59,60].

**6. Data analysis** encompasses collection, cleaning, statistical analysis, visualization, and validation. GPAI systems increasingly automate much of the analysis pipeline, from preprocessing to statistical modeling and interactive exploratory analysis[47,61,62]. Modern systems collect and integrate diverse data types and automate data cleaning and preprocessing, with high agreement with expert analyses within end-to-end automated workflows[22,25,42,63].

**7. Results interpretation** involves synthesizing experimental findings, evaluating hypotheses against results, and contextualizing findings within broader scientific knowledge. GPAI systems can now integrate computational and experimental data to connect predictions with outcomes and employ multi-agent approaches to critically evaluate results[22,25,26]. Advanced systems can compare experimental results to hypotheses and contextualize findings within existing knowledge[27,42].

**8. Manuscript preparation** includes documenting methods, presenting results, managing references, creating figures and tables, facilitating data sharing, and handling writing and revision. GPAI systems can now generate complete manuscript drafts with methodological descriptions and iteratively improve them through automated feedback[43,64]. Modern systems document experimental methods, manage references, generate figures, and prepare shareable data and code[24,27,42,47,65].

**9. Publication process** involves journal selection, submission, screening, peer review, and revision. GPAI assistance is emerging in areas such as journal and reviewer recommendation, formatting for submission, and generating peer reviews[66–69]. GPAI-generated feedback approximates human reviewers, and editorial experiments have found AI reviews sufficiently accurate to be helpful; GPAI can identify several (though not all) mathematical and conceptual errors[70–72].

Each of these main tasks comprises several subtasks with varying potential for GPAI acceleration. While these main tasks usually proceed in the order presented, research projects often require iterative transitions between tasks, such as when feedback from reviewers requires a return to experimental design or when unexpected results necessitate revisiting hypothesis generation. Notably, experiment execution represents the physical dimension of research, while all other tasks are primarily cognitive. This framework forms the basis for our investigation of how different levels of GPAI capabilities can accelerate the biomedical research process.

*Integration of GPAI capability and research task frameworks*

Combining the frameworks introduced above yields a matrix of research acceleration scenarios across GPAI capabilities and major research tasks (Table 1).

This integrated framework allows us to systematically analyze how different levels of GPAI capability transform the research process, identify where the greatest potential for acceleration exists, and what bottlenecks might remain even with advanced GPAI systems.

*Table 1: Major research tasks vs. capability levels*

| Research task | No GPAI | Next-level GPAI | Maximum-level GPAI |
|---|---|---|---|
| Knowledge synthesis | Humans integrate all findings manually into cohesive frameworks. | GPAI merges intermediate findings, but humans lead the broader synthesis and narrative formation. | GPAI synthesizes multi-source data into frameworks, humans coordinate alignment to strategic goals or ethical standards. |
| Idea & hypothesis generation | All concepts come solely from human insight. | GPAI offers derivative ideas, suggestions or refinements; humans make the primary decisions. | GPAI proposes mostly novel, creative hypotheses with minimal human oversight; final acceptance may still require a human check. |
| Experiment design | All protocols are devised manually by humans. | GPAI suggests minor modifications to standard designs; major decisions require human approval. | GPAI autonomously creates optimized designs; humans primarily provide high-level guidance or ethical oversight. |
| Ethics approval & permits | Humans manually complete all documentation and navigate approval processes. | GPAI assists with form completion and provides guidance; humans manage key interactions and decisions. | GPAI autonomously prepares documentation and predicts approval requirements; humans verify and provide final sign-off. |
| Experiment execution | Humans conduct all procedures; no GPAI-driven instruments beyond basic machinery. | GPAI helps automate some repetitive tasks under human supervision; humans manage complex or critical steps. | GPAI-managed robotics carries out most tasks independently, with humans intervening only for specialized or ethical considerations. |
| Data analysis | Humans perform all data processing and analysis. | GPAI conducts intermediate-level analyses (e.g., trend identification); humans perform deeper contextual interpretation. | GPAI completes analyses, often revealing insights humans might overlook; final interpretation may require minimal human input. |
| Results interpretation | Humans derive and contextualize all conclusions. | GPAI highlights interesting patterns and generates preliminary conclusions; humans interpret the results. | GPAI formulates detailed interpretations with minimal human support; high-level human review remains a quality check. |
| Manuscript preparation | Humans write, organize, and edit manuscripts by hand. | GPAI helps draft manuscript sections (e.g., methodology) in close interaction with humans; GPAI refines language; humans structure the core narrative. | GPAI composes complete manuscripts, with humans offering only selective edits or high-level guidance regarding focus and style. |
| Publication process | Humans handle submissions, conduct peer review, and manage revisions entirely. | GPAI helps with formatting, requirement checks, and review response suggestions; humans conduct peer review and remain the main contact for publishers. | Separate GPAI systems autonomously manage the submission and review processes, handling all responses and revisions; humans review and give final approval. |

## Evidence for acceleration potential

To obtain concrete GPAI acceleration factors, we conducted a scoping review of literature on GPAI accelerating research. It reveals significant variation in potential speedups between cognitive and physical tasks, with cognitive tasks generally showing higher acceleration factors[22,27,48,59]. The evidence ranges from modest improvements to dramatic transformations in research timelines.

*Acceleration of cognitive tasks*

The strongest empirical evidence for cognitive task acceleration comes from recent applications of GPAI in research environments, though the measurement approaches and reported metrics vary across studies. For instance, recent industry reports estimate GPAI-driven R&D in drug discovery from initial research to the preclinical stage, yielding a **1.3-2x efficiency increase**, corresponding to a 25-50% reduction in cost and time[73]. In more specialized research contexts, GPAI has enabled a **2-4x speedup** for analysis of flow cytometry data, by reducing the required time from 10-20 minutes to 5 minutes, while maintaining expert-level accuracy[61].

Looking at related tasks outside the core of science, studies have reported a **1.3x speedup** in consulting tasks[74], a **1.7x speedup** (40% time decrease) in professional writing tasks[75], and a **2.2x speedup** (55% time decrease) for coding tasks, according to GitHub's internal report on Copilot[76].

The reported acceleration potential increases significantly with more advanced GPAI systems and setups optimized for automation. For example, in scientific knowledge synthesis, PaperQA2 demonstrated an **~75-300x speedup**, writing high-quality Wikipedia-style articles in 8 minutes[27] (a task that human editors report taking 10-40 hours[77]). In bioinformatics, a GPAI agent capable of fully automated multi-omic analyses reportedly just required 5 minutes for the exemplary task of identifying differentially expressed genes between bulk RNA-seq samples[47]. Compared to the 10-12 hours[78,79] reported by two bioinformatics facilities for the same type of task, this represents an approximate speedup of **~120-140x**.

Extending to complete research cycles, a GPAI agent team claimed to have developed SARS-CoV-2 nanobodies in a fraction of the time human researchers would have needed[25,80]. Another automated biomedical GPAI system called BioResearcher reported achieving a **~150-300x speedup** by completing full dry lab research cycles, from literature searches to the execution of computational experiments, in approximately 8 hours versus the traditional 7-14 weeks[22]. Similarly, the 'AI Scientist' demonstrated high efficiency through full automation in computer science research, capable of exploring research ideas in ~15 minutes (a rate of ~50 ideas in 12 hours)[24].

*Acceleration of physical tasks*

In laboratory settings, physical task acceleration also shows promising results, though generally with somewhat lower acceleration factors than purely cognitive tasks. An integrated robotic chemistry system achieved a **1.7x speedup** in synthesizing nerve-targeting agents compared to manual methods, completing the entire 20-compound library in 72 hours instead of 120 hours, reportedly with comparable quality[81].

Protein engineering with integrated GPAI testing and feedback demonstrated a reduction in project duration from 6-12 to six months **(1-2x speedup)** in real-life testing (including shipping delays). The authors suggest that with better planning it could be reduced to two months **(3-6x speedup)** and in the best-case scenario of continuous operation to just 1-2 weeks **(~15-50x speedup)**[48]. A GPAI-driven microbial culturomics platform overcomes the variability of manual methods by using imaging to autonomously inform colony selection, yielding a more than **20x speedup** by achieving an isolation throughput of 2,000 colonies per hour in an integrated pipeline[59]. An automated materials discovery platform integrated ML screening, robotic synthesis, and characterization, reportedly reduced material sintering times from 2-6 hours to 10 minutes **(12-36x speedup)** and was noted to reduce entire processes from hours or days to minutes[82].

An automated chemical workflow handling 16 parallel samples conducted 688 experiments in 8 days. Compared to manual methods, which were estimated to take half a day per experiment, this represents a ~**40x speedup**. In addition to these results, the study reported estimated acceleration factors of ~**10x-100x** compared to conventional workflows, where the lower range corresponds to semi-automated methods and the higher end to manual approaches[28]. Self-driving labs that integrate robotics, additive manufacturing, and GPAI were projected to accelerate materials and molecular discovery by **10–100x** through combining gains from robotics (2x), active learning (5-20x), process intensification (up to 100x) and continuous operation (2-3x)[83].

Based on internal industry data, one prominent cloud lab suggests its GPAI implementation enables a **2x speedup** in time-to-publication (from an average of 1.96 years to one year) and claims to generate publication-quality data **90x faster** by reducing traditional 3-month timelines to 24 hours[84]. In a notable anecdote, a PhD student reported replicating years of their previous project's work in just one week using an automated robotics platform with 24/7 operation **(~100x speedup)**[85].

*Challenges and constraints of biomedical research acceleration*

The empirical evidence reviewed above highlights the potential for GPAI to accelerate both cognitive and physical research tasks, with some studies demonstrating order-of-magnitude improvements. These impressive figures often reflect optimized scenarios or specific sub-tasks. Realizing such acceleration consistently across the entire research lifecycle is subject to various **practical, biological, infrastructural, and institutional constraints.**

One major category of reported constraints relates to the **implementation and operation of automated systems and self-driving laboratories (SDLs)**. Studies note that creating robust SDLs requires significant investment and complex integration of automated experiments with GPAI decision-making[86,87]. Even once operational, researchers report challenges in automated or cloud lab environments including remote troubleshooting difficulties, reduced experimental flexibility for exploratory research, and limitations in applicability for academic settings characterized by frequent directional pivots[88]. Comparing the effectiveness of different SDLs is also reportedly difficult due to challenges in defining standardized performance metrics that capture the nuances of diverse lab setups[87]. Finally, translating discoveries made in controlled SDL environments to real-world applications faces hurdles related to storage stability, limited resources, and self-sufficiency without expert supervision[89].

Another set of limitations highlighted in the literature concerns data and the GPAI models themselves. The performance of models central to GPAI and SDLs can be hampered by shortcomings in available data, such as the common lack of negative results or detailed metadata in published literature[90]. Building robust GPAI decision-making models often requires large, high-quality, information-rich datasets, the generation of which can be a bottleneck[91].

Perhaps the most fundamental constraints reported are inherent biological and physical limits. While automation can speed up workflows like liquid handling or data acquisition, the underlying biological processes often have irreducible timescales[88]. Cell-based experiments, for instance, remain resource-intensive and subject to variability, with limits imposed by factors like maximum cell growth rates under specific conditions[91]. SDLs optimize experiments around these biological components rather than altering their intrinsic limits[86,90]. Even the speeds of biochemical processes, like enzyme kinetics or protein folding, impose natural limits that automation cannot bypass[89]. Finally the complexity of biology also presents a challenge, as fully understanding and predicting cellular behavior remains difficult without comprehensive perturbation data, even when advanced models are utilized[88,91].

Aside from the technical and biological hurdles, significant limitations arise from established social and institutional structures and processes. Although GPAI can accelerate certain tasks, the overall pace can be slowed by procedural delays inherent in the current academic system, as they may not scale readily with technological advancements. Ethics approval processes average between 50 and 138 days[92]. Similarly, publication faces substantial delays: preprint-to-publication averages 199 days[93], with submission-to-publication times within journals ranging from 91 to 639 days[94]. The peer review process itself was reported to take 17 weeks in one study[95], despite reviewers spending only about six hours per review in each round[96].

These documented limitations, which have been identified in the literature alongside the potential for acceleration, suggest that achieving maximum theoretical acceleration across entire research workflows poses significant practical, technological, and ethical challenges. In fact, even beyond these operational and inherent limitations, rigorously assessing the extent of GPAI-driven acceleration itself presents methodological complexities. The interpretation of reported accelerations requires clear baseline values and system boundaries. Highly task-specific accelerations, such as GPAI agents that rapidly design nanobodies[25] or automate the planning and coding of bioinformatics analyses[22,47], are valuable but must be distinguished from reductions in overall project duration.

Assessing the acceleration of research driven by GPAI requires methodological rigor, as the perceived benefits depend on the chosen benchmark (e.g., humans, optimized laboratories, or state-of-the-art automation). For example, robotic systems accelerate workflows such as chemical synthesis[28] or high-throughput screening[85] primarily through parallelism and continuous operation, rather than pure task speed. Such throughput gains are significant, but must be compared with appropriate advanced baselines—not just sequential execution by humans—and must take into account fixed setup costs that can reduce the benefit of switching to automated workflows, especially in isolated or small-scale projects.

Similarly, the acceleration offered by GPAI in the computer-aided discovery of therapeutics[31] or proteins[48] can only be judged when the validation effort for GPAI-generated hypotheses is taken into account. Therefore, precise reporting standards are crucial, requiring transparency regarding benchmarks, system boundaries, the distinction between actual speed and throughput, and full consideration of all operating and validation costs in order to accurately assess the benefit of GPAI.

## Acceleration factors with GPAI levels

Our literature review reveals a bimodal distribution of acceleration factors across research tasks, with most observed values clustering either at lower levels (below 3x) or higher levels (above 10x for physical tasks and above 50x for cognitive tasks) (Figure 3, Table S2). This pattern suggests two distinct regimes of reported GPAI-driven acceleration: incremental acceleration attainable today across the entire research process (Next-level GPAI), contrasted with transformative acceleration, achievable only through advanced systems and configurations (Maximum-level

GPAI), and currently restricted to specific research tasks—thereby directing attention toward the theoretical maximum of research acceleration.

To translate these empirical findings into practical modeling scenarios, we assigned two distinct acceleration profiles to the capability levels defined in our framework:

1. **Next-level GPAI:** This profile models the **current acceleration potential** of current GPAI systems as they diffuse through the research ecosystem. Based on the lower range of our empirical findings, we estimate acceleration factors of **2× for both cognitive and physical tasks**. We assume that these represent realistic, immediately achievable improvements that organizations can expect when implementing current GPAI and lab automation technologies, as already demonstrated in preclinical drug discovery for pulmonary fibrosis[31].

2. **Maximum-level GPAI:** This profile explores the **transformative acceleration potential** of research with future, highly advanced GPAI systems. Drawing from the upper ranges of documented capabilities, we estimate acceleration factors of **100× for cognitive tasks** and **25× for physical tasks**—values deliberately selected to err on the conservative side of reported factors (Figure 3). While these factors may seem extraordinary, they represent acceleration potentials that have been demonstrated in specific contexts and thus provide evidence for possible future scenarios.

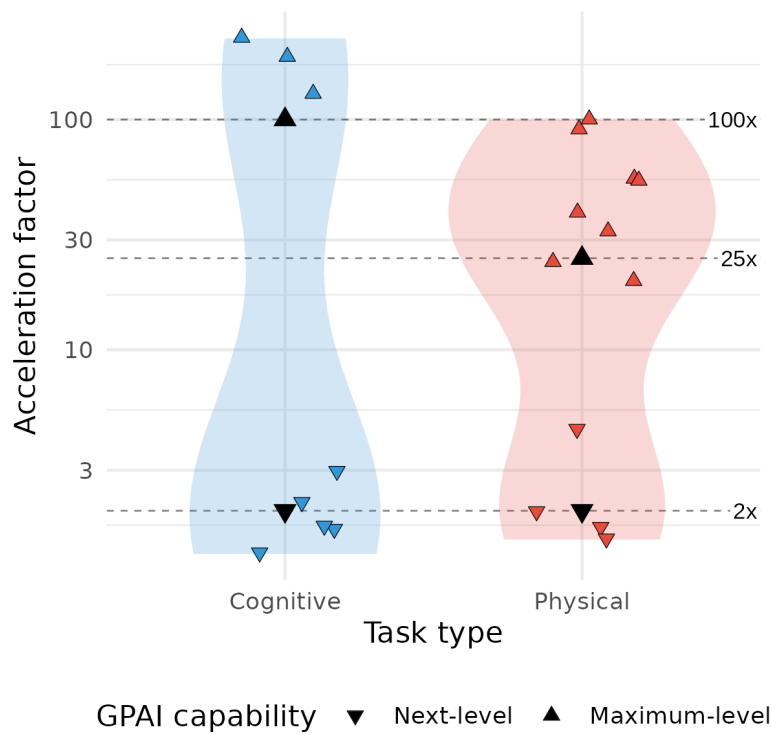

*Figure 3: Reported acceleration factors for cognitive and physical research tasks.* Violin plots depict the distribution of 20 acceleration factors on a $\log_{10}$-scale extracted from 16 publications for cognitive (blue) and physical (red) tasks. Individual studies are shown as jittered triangles; downward symbols (▼) indicate next-level capability, upward symbols (▲) indicate maximum-level capability. Four black triangles and three dashed lines (at 2×, 25×, 100×) denote the factors chosen for our modeling scenarios.

*Biological time constants in acceleration modeling*

It is important to note that the acceleration factors cited above derive primarily from high-throughput in-vitro experiments and computational tasks. However, biomedical research, particularly involving whole organisms, contains certain irreducible processes that cannot be accelerated beyond natural biological limits (such as time for cell growth, animal model development, or tumor progression). We therefore include a "non-compressible" time constant in our model representing irreducible intervals dictated by biological processes that remain fixed regardless of technological advancement.

## Acceleration scenario modeling

Taken together, we propose the following simple formula to estimate research time:

**Total research time = (Compressible time ÷ Acceleration factor) + Non-compressible time**

To demonstrate the effect of different acceleration scenarios on research time, we model a hypothetical 3-year biomedical research project representative of a typical PhD project duration. In this example we assume 24 months of cognitive work and 12 months of physical experimental work, of which 3 months represent biological time constants that cannot be compressed (Table 2).

*Table 2: Worked example of GPAI-driven reduction in project duration for a hypothetical 36-month research project with a 3-month biological non-compressible time constant.*

| Project duration (in months) | Physical: No GPAI (9+3= 12 months) | Physical: Next-level GPAI 2x-acceleration (9/2 + 3= 7.5 months) | Physical: Maximum-level GPAI 25x-acceleration (9/25 + 3= ~3.4 months) |
|---|---|---|---|
| Cognitive: No GPAI (24 months) | 36.0 (=1x) | 31.5 (~1.1x) | 27.4 (~1.3x) |
| Cognitive: Next-level GPAI 2x-acceleration (24/2 = 12 months) | 24.0 (=1.5x) | 19.5 (~1.8x) | 15.4 (~2.3x) |
| Cognitive: Maximum-level GPAI 100x-acceleration (24/100 = ~0.2 months) | 12.2 (~3x) | 7.7 (~4.7x) | 3.6 (=10x) |

Cognitive tasks offer greater potential for GPAI-driven acceleration due to their longer initial durations and higher acceleration factors, lacking the inherent limitations of biological processes and requiring less infrastructural investment. Still, achieving the most significant acceleration across biomedical research depends on maximizing GPAI capabilities in both cognitive and physical domains.

Applying maximum-level GPAI to both cognitive and physical tasks in our example reduces a 3-year (36-month) biomedical research project to 3.6 months, a 10x acceleration despite the biological time constant. While this constant constitutes a small fraction of the total timeline without GPAI (3 of 36 months ~8.3%), it becomes the dominant factor (3 of 3.6 months ~83%) under maximum-level acceleration.

Fields heavily dependent on in-vitro or computational approaches may realize acceleration factors approaching our maximum estimates, while those requiring extensive in-vivo work will experience more modest overall timeline reductions due to the presence of larger biological time constants.

### Exploratory expert elicitation

In order to assess both the plausibility of our acceleration factors and limiting conditions, as well as the prevailing attitude and expectations of biomedical researchers toward GPAI for accelerating research, we conducted an exploratory elicitation with eight biomedical expert researchers. In our structured survey, they reflected on how our findings would apply to projects they had led from conception to publication in high-impact journals. The experts reported an average project duration of 72 months (Figure 4), which is twice as long as our hypothetical example, but is consistent in the proportional distribution between cognitive and physical tasks: cognitive tasks took up 73% of project time (52 months), which is similar to the 67% in our hypothetical project. Experts identified "experiment execution," "publication process," and "data analysis" as the most time-consuming research tasks, while "ethics approvals and permits" and "knowledge synthesis" were rated as the least time-consuming.

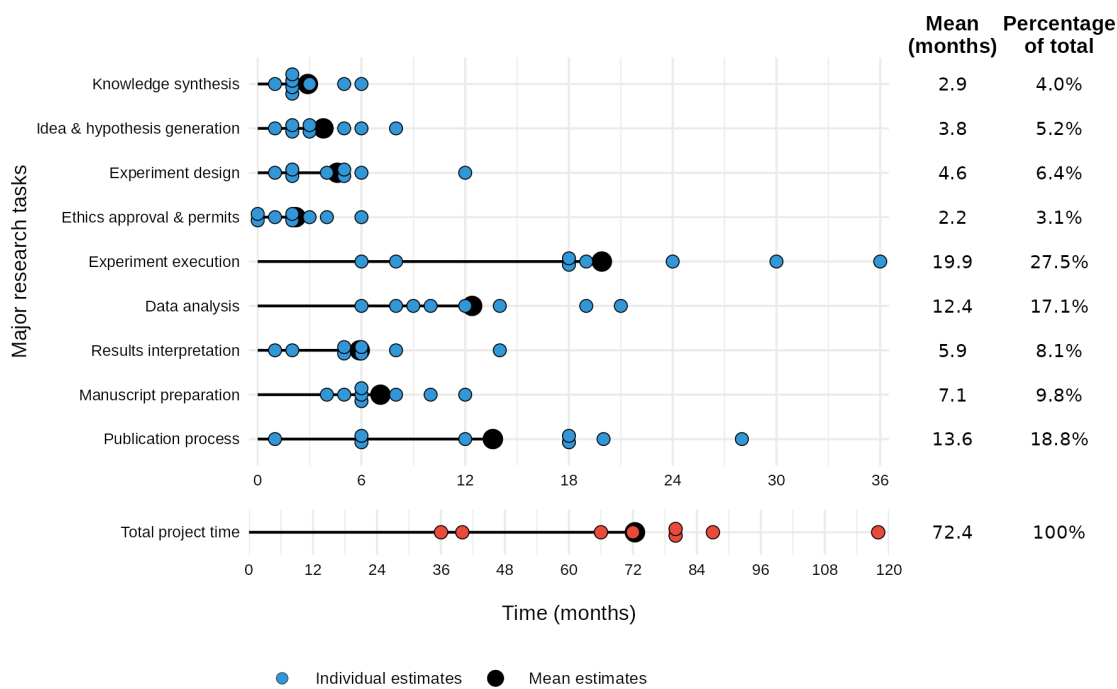

*Figure 4: Timeline of major research task durations across studies.* Horizontal dots show individual estimates (one dot per study) of time spent (in months, on a 6-month–interval axis) on each major research task, with lollipop-plots indicating the task's mean. On the right, for each task, the rounded mean duration (months) and percentage of the total project time is shown. Below, a separate row presents individual and mean estimates for overall project duration (in months, on a 12-month–interval axis).

When **evaluating our estimates for maximum-level acceleration** (~100-fold cognitive, ~25-fold physical), biomedical experts judged experiment design and execution, and hypothesis generation to be strongly overestimated, while greater acceleration potential was deemed plausible for administrative tasks. Respondents consistently considered our acceleration estimates overestimated for experimental design (7/7 responses), experimental execution (7/8) and hypothesis generation (7/8). In contrast, experts considered high acceleration factors plausible for structured administrative processes: ethics approval (4/7), manuscript preparation (4/8), and publication processes (3/8), with the rest of the responses mixed between over- and underestimation (Figure 5).

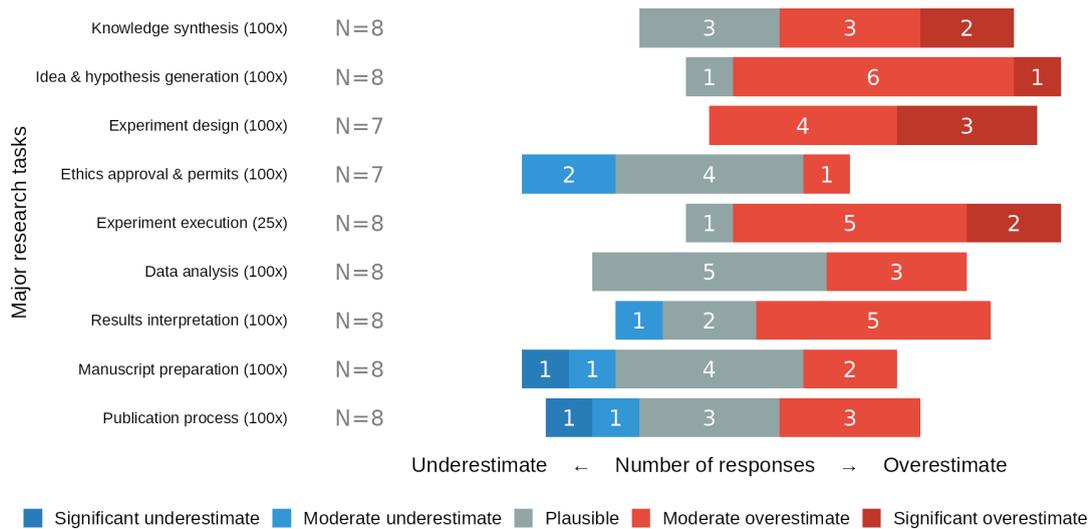

*Figure 5: Perceived plausibility of maximum-level GPAI-acceleration factors across the nine major research tasks. Colors denote responses: significant underestimate (dark blue), moderate underestimate (light blue), plausible (grey), moderate overestimate (light red), and significant overestimate (dark red), with number of responses in white.*

We asked experts to rate the **significance of various potential bottlenecks** (Figure 6). While many factors showed a mixed response, scientific community assimilation was rated by all respondents as a moderate (2/8 responses), major (4/8) or crucial limit (2/8). In contrast, human strategic direction was seen as a lesser constraint, with a majority rating it as a minor (4/8) or insignificant limit. There was a marked consensus that stakeholder coordination is only a moderate limit (5/8).

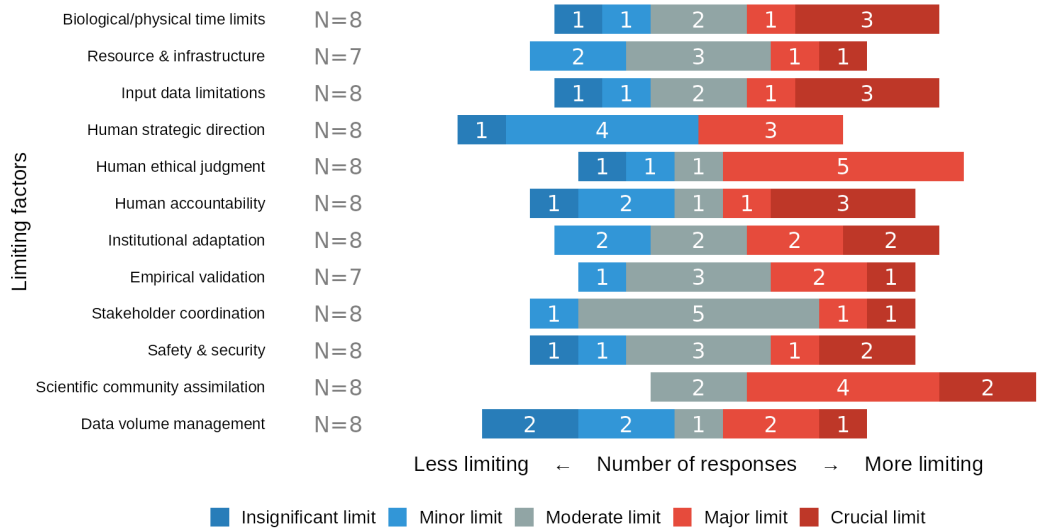

*Figure 6. Perceived severity of factors that may limit GPAI-driven research acceleration.* Colors encode response categories: insignificant limit (dark blue), minor limit (light blue), moderate limit (grey), major limit (light red), and crucial limit (dark red), with number of responses in white.

In addition to the quantitative ratings, **experts provided general considerations**, in which they highlighted irreducible biological and social constraints. One researcher noted that for their project, "the blood sampling of 200 individuals simply take[s] a definite time," while another pointed to the "fundamental time-frame of the experiment (i.e., looking at a 3 month effect after intervention)." The limitations of social processes were also stressed, "the speed of publication with peer review and also the response of the co-authors cannot be changed," underscoring that institutional adaptation and human coordination remain important bottlenecks. Experts also emphasized practical challenges in system integration and the socio-economic barriers to adoption. One expert noted the difficulty of "interfacing of various output/input systems." and also pointed to the "Cost/Benefit ratio," suggesting that the high upfront cost requires "phenomenal trust in results" (see Tables S3–S6 for full survey results and Information S7 for survey interface).

# Discussion

Our investigation suggests that the availability and diffusion of highly capable GPAI systems might cause a fundamental transformation in how biomedical research will be conducted. The results of the exploratory expert elicitation support our research project life cycle's time distribution between cognitive and physical tasks (73% versus 67% for cognitive tasks). While the experts expressed skepticism about extreme acceleration of experiment planning and execution, as well as hypothesis generation—suggesting that transformative research activities will continue to be constrained by human judgment and biological time constants—they considered high acceleration factors to be more plausible for structured processes such as manuscript preparation and the time-intensive process of publication. Most significant, however, is the unanimous concern about the scientific community's ability to assimilate, suggesting that this could ultimately limit the pace of scientific progress.

## Limits to acceleration

While GPAI capabilities continue to advance rapidly, several key factors may limit the translation into overall research acceleration:

**Technical and infrastructural limitations** encompass the technological prerequisites for GPAI-accelerated research and the shift from local to centralized cloud labs. These prerequisites include the availability, capacity and flexibility of cloud labs and automated devices, computing resources for training and operating GPAI systems, the capital expenditure and high fixed costs of self-driving labs, and the quality and accessibility of research data.

**Biological and physical limitations** impose fundamental boundaries on acceleration. Cell division rates, organism development cycles, or the duration of clinical trials follow natural timescales that cannot be compressed beyond certain limits. Similarly, material handling and physical operations have speed limitations due to physical limits and safety considerations. However, growing effectiveness of in-silico simulation may affect the balance between fast computational experiments and slow wet-lab experiments.

**Institutional and regulatory factors** pose significant potential barriers to acceleration. Social processes, including ethical review, peer review, and publication, present particular challenges as they involve human judgment, predefined institutional procedures, and regulatory requirements that may not readily adapt to technological advances. The pace of GPAI adoption is significantly determined by the adaptability of institutions and workforces.

**Human oversight requirements** may remain indispensable for some aspects of research. Strategic direction continues to be determined primarily by humans, and ethical considerations require human review of certain decisions. Quality control often requires human expertise. The effectiveness of collaboration between humans and GPAI depends on the trust researchers have in GPAI outputs, and more broadly, the scientific community's ability to assimilate and adapt to these new tools could ultimately limit the pace of scientific progress.

In light of these potential limiting factors, the full potential of GPAI-driven acceleration may only be realized when technological change is complemented by transformations of social processes, infrastructure, and governance frameworks.

## Transformation of research processes

**Review and publication processes** like peer review and ethics approval may face strong incentives to streamline and transform. Our study suggests that publication processes constitute a substantial fraction of the overall project duration, and can be plausibly accelerated in a very substantive manner. Advanced GPAI systems could prove more effective than humans at aspects of reviewing, potentially checking protocol compliance and ethical considerations with greater reliability and comprehensiveness. This could shift review processes from sequential, time-intensive review cycles—whose duration stems from coordination challenges—towards more continuous and immediate monitoring. Consequently, human roles may shift from providing detailed reviews to general reflection, setting high-level standards, strategic oversight of GPAI, and handling complex cases, potentially requiring specialized skills for GPAI-human collaboration.

GPAI has various potential applications within the peer review process, from submission preparation and reviewer-paper matching to providing direct assistance with evaluation and formulating clear reviews[68]. GPAI feedback can supplement the scientific process, particularly during early manuscript development, though human review currently needs to remain the foundation of the review process[71]. Though GPAI can reduce reviewer burden in human-in-the-loop settings for many but not all cases, implementation risks like bias and potential misuse demand systematic study[72].

**Metrics of scientific quality** may need revision, and the optimal balance between speed and thoroughness requires careful consideration. Existing incentives valuing perceived novelty and quantity over reproducibility could exacerbate the proliferation of unreliable findings, but GPAI might also enable the highly detailed and transparent documentation needed for reproducibility.

The **dynamics of goal-setting and exploration** in scientific inquiry may change fundamentally. Unlike human teams requiring lengthy onboarding, GPAI systems can be deployed immediately, reconfigured, or scaled to explore hypotheses without administrative delays, allowing a rapid switch of research directions. GPAI systems could bridge disciplinary boundaries by processing vast amounts of literature and facilitating novel connections between previously separate bodies of knowledge. In such a setting, steering research directions requires attention to potential biases in dominant GPAI models, which may influence which research areas receive attention. GPAI could lead to a homogenization of research approaches, which undermines the diversity of perspectives that has historically driven scientific innovation. Mechanisms to identify such biases and incentivize diverse approaches to problem-solving may be needed.

**Resource requirements** will become increasingly important. Infrastructure capacity, material and energy availability for continuous GPAI operations, and computing power and data access may emerge as critical limiting factors. Institutions unable to adapt to these new requirements may face significant competitive disadvantages.

## Policy implications

GPAI-driven acceleration requires significant policy challenges, from resource allocation and workforce adaptation to safety and governance.

**Resource allocation.** The extent to which different fields and institutions benefit from GPAI will depend on their access to capital, computational infrastructure, laboratory automation and frameworks for effective human-GPAI collaboration. GPAI-driven research acceleration risks increasing inequalities between institutions and political entities, which can lead to a concentration of progress and significant power imbalances. The resulting capture of intellectual property and market share could further entrench these inequalities, even if some institutions remain only slightly behind the cutting edge of GPAI-driven progress. To counteract these risks, policymakers should proactively fund GPAI-driven research in two complementary areas: building self-driving laboratories that can flexibly execute diverse physical experimental tasks, and expanding access to frontier model capabilities and computational resources required for cognitive research tasks.

**Workforce adaptation.** GPAI may complement human researchers by shifting their focus to higher-level strategic direction, analysis and validation. Research organisations may therefore need to restructure to take full advantage of GPAI opportunities. This includes providing GPAI knowledge and skills to current and future scientists, including overseeing GPAI projects and critically interpreting their results, as well as updating incentive mechanisms to reward effective and transparent use of GPAI.

**Preventing misuse.** Care must be taken when applying GPAI to research on dangerous biological agents to avoid increasing dual-use concerns. Effective international coordination is essential, including common approaches to monitoring, harmonised standards, uniform transparency requirements and uniform ethical guidelines for both advanced GPAI and biological laboratory automation.

**Efficient governance.** Traditional governance structures struggle to keep pace with GPAI-accelerated science. This requires efficient frameworks that allow for continuous monitoring and rapid adjustments so that policies can be iteratively updated as new risks and opportunities emerge. Such frameworks may require updating research assessment methodologies and incentive structures to maintain scientific quality and to achieve best possible outcomes. Careful integration of AI tools into the governance processes themselves may help meet this challenge.

## Limitations of this study

Our study has several important limitations. The generalizability of our framework remains limited by significant variations in institutional structures and research practices across scientific fields and organizations. Biomedical subfields possess distinct characteristics that may lead to uneven acceleration potentials not fully captured in our estimates.

While our study focused on estimating the potential acceleration of established processes and research tasks, long-term transformation by GPAI would likely go beyond mere acceleration of existing processes and introduce new paradigms in research. In our worked example, once cognitive and physical tasks are maximally accelerated, the irreducible time constants inherent to lab experiments become the dominant bottleneck (3 out of 3.6 months). Paradigm-shifting strategies that we excluded—such as replacing in vivo/vitro experiments with in silico experiments or introducing AI-first prioritization to decrease the amount of required experiments—could reduce this remaining constraint and create a qualitatively new research dynamic. Assessing the feasibility, risks, and governance implications of such transformations is beyond the scope of our investigation. This boundary condition limits the gains we report but increases the robustness of our estimates by restricting them to accelerations that are evidence-based, plausible and relevant within the current research paradigm.

Our framework of major research tasks categorized into discrete cognitive and physical tasks, while providing analytical clarity, represents necessary simplifications that may obscure important research acceleration dynamics. Many biomedical research tasks are inherently hybrid or have intertwined components, with real research characterized by frequent iterative switching between processes rather than discrete, sequential phases. Additionally, while our sequential framework does not capture real-world task parallelization by multiple team members or project downtimes and delays, we believe that these opposing influences cancel each other out to a certain extent, so that the resulting inaccuracy remains low.

The projection of GPAI capabilities inherently involves substantial uncertainty. The interactions between accelerated research processes, institutional structures, and social systems introduce additional unpredictability. Our modeling approach necessarily simplifies these dynamics and does not account for feedback loops or emergent phenomena that could significantly influence real-world outcomes.

Current empirical evidence provides an incomplete foundation for robust predictions. While promising examples of GPAI-accelerated research are emerging, long-term and large-scale implementations are still scarce. Our literature review revealed considerable heterogeneity in reported acceleration factors and how they were determined. The adaptation rate of social and institutional processes to technological acceleration represents another significant unknown with limited historical precedent.

The expert elicitation introduces potential selection bias and the number of experts consulted was relatively low at eight individuals, which limits the robustness of the results. Nevertheless, this elicitation provides an initial assessment of the plausibility of our key assumptions and findings, and reveals the prevailing attitude of experts toward GPAI and its potential to accelerate research.

Despite these constraints, our multi-method approach establishes a foundation for elucidating the limits to biomedical research acceleration. While our current estimates contain high uncertainty, they do suggest that order-of-magnitude acceleration through GPAI implementation is plausible. By identifying key drivers and barriers to acceleration, this work offers valuable guidance for scenario planning and policy development in an increasingly GPAI-enabled research landscape.

## Acknowledgements

We thank the experts who participated in our expert elicitation, providing insights into GPAI's potential impact on research timelines by estimating the actual time spent on the outlined major research tasks within a published project and assessing both the potential for acceleration and its limitations: Nikolaus Fortelny, Máté G. Kiss, Bernhard Kratzer, Thomas Krausgruber, Anna Redl, Rob ter Horst, Peter Traxler, Dimitrios Tsiantoulas.

## Methods

To analyze the potential acceleration of biomedical research through GPAI, we employed a multi-faceted approach combining framework analysis, literature review, and expert feedback. This methodology enabled us to systematically evaluate both current capabilities and future potential while maintaining practical relevance.

**Framework development**: We began by analyzing existing conceptual frameworks related to GPAI capabilities and research automation. We identified and reviewed four relevant frameworks: DeepMind's "Levels of AGI," which differentiates between narrow and broad AI systems[33]; SAE's "Driving automation systems," which describes levels of interaction between humans and increasingly autonomous vehicles[34]; "AI agents for biomedical discovery," which details increasing levels of AI agency in biomedical research[35]; and "Self-Driving Laboratories," which maps the integration of software and hardware processes in laboratories[36]. Based on these frameworks and the domain-specific requirements of biomedical research, we developed a simple unified framework focusing on cognitive and physical capabilities across three key levels.

Drawing on these frameworks and analyzing the specific requirements of biomedical research, we synthesized a **simple, unified framework of GPAI research capability** with two key dimensions: **cognitive capability** and **physical capability**. Cognitive capability encompasses research activities primarily involving information processing, analysis, and decision-making, including literature review, hypothesis generation, experimental design, data analysis, result interpretation, and manuscript preparation. Physical capability involves laboratory procedures, experimental setup, and material handling through robotics, lab automation, and experiment setup through experiment execution.

**Research process mapping**: We systematically analyzed the biomedical research lifecycle to identify major research tasks and their subcomponents. We mapped the complete research process from initial knowledge synthesis to final publication, identifying nine major research tasks encompassing the full research cycle. We then decomposed each major task into constituent subtasks with distinct characteristics and classified tasks according to their primary capability dimension (cognitive vs. physical). This structured mapping provided a foundation for applying our capability framework and assessing acceleration potential across the research process.

**Scoping literature review**: We conducted a scoping review of literature on GPAI-driven research acceleration published within the last five years up to March 2025, to reflect current technology. Our search strategy identified relevant studies using keywords related to GPAI, research automation, and acceleration across major scientific databases. Studies were included if they reported concrete acceleration metrics for research tasks, were published in peer-reviewed journals or reputable preprint servers, focused on biomedical research or closely related fields. Because only a handful of studies report quantitative speed-up, a classical systematic review (fixed protocol, database preregistration) was not feasible.

For each included study, we extracted the research tasks addressed, the GPAI technology or methodology employed, and quantitative acceleration metrics (e.g., time reduction, throughput increase). When studies reported acceleration factors as ranges, these were converted to point estimates using the arithmetic mean (e.g., a range of [150–300] becomes 225; see Table S2) to facilitate the quantitative analysis presented in Figure 3. One paper reports 10x reductions for semiautomated methods and 1000x reductions in researcher time for manual methods. We report more conservative 10-100x reductions because we account for total experiment duration, not just researcher time, yielding approximately 100x overall acceleration (1000 experiments require 500 days manually versus 5.5 days autonomously—5 days experimental time + 0.5 days researcher time).[28]. We then mapped the findings to our framework, identifying task-specific acceleration potential, patterns across cognitive and physical dimensions, and current vs. theoretical acceleration limits. This approach allowed us to ground our acceleration estimates in empirical evidence while acknowledging uncertainty in projections of maximum capability levels, interpreting reported values as current upper limits while accounting for potential publication bias favoring optimistic estimates.

## Framework integration

We integrated findings from the literature review to develop our unified acceleration framework. We mapped acceleration metrics to specific research tasks and capability levels, complemented our findings using real-world research timeline examples, synthesized results into a concrete framework showing potential acceleration across different scenarios, and identified key bottlenecks and rate-limiting factors that might constrain overall research acceleration. This integrated approach ensured our framework was both theoretically sound and practically relevant, reflecting the current state of the literature while acknowledging the limitations of existing evidence.

## Expert elicitation

To complement our literature findings, we developed a structured questionnaire (see Information S7 for original survey interface) aimed at researchers with direct experience managing biomedical research projects. The elicitation process, deployed via the web-based platform Alchemer, was designed to capture real-world experience and professional judgment regarding potential research acceleration through advanced GPAI systems and their limitations. Given the novelty of the effects of GPAI, our goal was not to achieve statistical generalizability, but rather to capture the current sentiment and informed judgment of domain experts on the identified potential and limitations of GPAI in accelerating research.

### *Expert selection*

We identified and contacted authors who published biomedical studies in high-impact journals (3x Immunity, 2x Nature, 1x Nature Genetics, 1x Allergy, 1x Cell Systems) between 2020 and 2025. Selection criteria ensured participants had led (as first authors) or coordinated (as last authors) the published research project. This approach targeted researchers with experience across the full research lifecycle from conception to publication.

### *Elicitation protocol and data analysis*

To provide context and calibration, respondents were presented with a table summarizing acceleration levels reported in recent literature for each major research task, including relevant citations. This helped ensure estimates were grounded in current technological capabilities while allowing for informed projection of future potential. Building on this context, we structured our survey in four main sections:

1. **Project timeline estimation**: Respondents provided the total duration (in months) of a specific biomedical research project they had led or coordinated. They then estimated how time was allocated across the nine major research tasks we identified in our framework. This established a baseline of actual research timelines against which to assess potential acceleration. We calculated mean project durations and mean time allocation for each of the nine research tasks, then computed the proportion of time spent on cognitive versus physical tasks.

2. **Plausibility of acceleration estimation**: We presented a hypothetical future scenario defined by the availability and universal use of maximum-level GPAI - defined as future GPAI with autonomous decision-making, advanced multi-disciplinary reasoning, and deep integration with robotics. Based on our scoping review of recent literature, we provided respondents with estimated accelerations of ~100x for primarily cognitive tasks and ~25x for primarily physical tasks. For each of the defined nine major research tasks, respondents evaluated the plausibility of achieving these pre-defined acceleration factors in the context of their own project. Responses were collected on a five-point Likert scale ranging from "Significant Overestimate" to "Significant Underestimate". Response frequencies were tabulated for each research task across the five-point scale, and we reported the number of respondents who rated each acceleration estimate as plausible, overestimated, or underestimated.

3. **Evaluation of limiting factors:** To identify key bottlenecks that might prevent achieving theoretical acceleration maximums, respondents were presented with a list of 12 literature-derived potential limiting factors. These spanned technical constraints (e.g., fundamental biological time limits), resource constraints (e.g., energy, infrastructure), and human/institutional factors (e.g., ethical oversight, regulatory adaptation). Respondents rated the significance of each factor in limiting practical acceleration on a five-point Likert scale from "Insignificant Limit" to "Crucial Limit". We computed the distribution of ratings for each of the 12 limiting factors and identified factors with a unanimous or near-unanimous consensus among respondents.

4. **General considerations**: An open-ended section allowed respondents to share additional thoughts on research acceleration, potential risks, and ideas for mitigating bottlenecks. The responses were thematically analyzed to identify recurring themes related to implementation challenges, constraints, and opportunities for research acceleration.

## Author contributions

K.H., C.C., S.R., and M.S. conceptualized the study and developed the methodology. K.H., C.C., and S.R. performed the formal analysis and investigation and prepared the original draft. S.H. contributed to the investigation and methodology development. C.B. provided critical feedback on the manuscript. M.S. supervised and administered the project.

During manuscript preparation, authors used AI models from the providers OpenAI, Google, and Anthropic to refine language and assist with literature summarization, and the services from Grammarly and DeepL for grammar and style checking. All content generated using these tools was carefully reviewed, edited, and verified by the authors, who assume full responsibility for the final manuscript.

## Data availability

All anonymized elicitation data, analysis scripts and figure-generation code are available at GitHub: https://github.com/OpenBioLink/ResearchAcceleration

# References

1. OECD. *Artificial intelligence in science: challenges, opportunities and the future of research*. (OECD, 2023). doi:10.1787/a8d820bd-en.
2. Solow, R. M. Technical change and the aggregate production function. *Rev. Econ. Stat.* **39**, 312 (1957).
3. Agrawal, A. K., Gans, J. S. & Goldfarb, A. *The economics of artificial intelligence: an agenda*. (University of Chicago Press, 2019). doi:10.7208/chicago/9780226613475.001.0001.
4. World Economic Forum. Top 10 Emerging Technologies of 2024. https://www.weforum.org/publications/top-10-emerging-technologies-2024/ (2024).
5. Jumper, J. *et al*. Highly accurate protein structure prediction with AlphaFold. *Nature* **596**, 583–589 (2021).
6. EU-AI Act. Article 3: Definitions | EU Artificial Intelligence Act. https://artificialintelligenceact.eu/article/3/ (2025).
7. Fortunato, S. *et al*. Science of science. *Science* **359**, (2018).
8. Wu, L., Wang, D. & Evans, J. A. Large teams develop and small teams disrupt science and technology. *Nature* **566**, 378–382 (2019).
9. Rao, H. *et al*. How many authors does it take to publish a high profile or classic paper? *Mol. Biol. Cell* **33**, (2022).
10. Brunson, J. C., Wang, X. & Laubenbacher, R. C. Effects of research complexity and competition on the incidence and growth of coauthorship in biomedicine. *PLoS ONE* **12**, e0173444 (2017).
11. Chu, J. S. G. & Evans, J. A. Slowed canonical progress in large fields of science. *Proc Natl Acad Sci USA* **118**, (2021).
12. M-A-P Team. SuperGPQA: Scaling LLM Evaluation across 285 Graduate Disciplines. https://arxiv.org/abs/2502.14739 (2025).
13. Quan, S. *et al*. CodeElo: Benchmarking Competition-level Code Generation of LLMs with Human-comparable Elo Ratings. *arXiv* (2025) doi:10.48550/arxiv.2501.01257.
14. EpochAI. AI Benchmarking Dashboard | Epoch AI. https://epoch.ai/data/ai-benchmarking-dashboard (2025).
15. Artetxe, M. *et al*. Efficient Large Scale Language Modeling with Mixtures of Experts. *arXiv* (2021) doi:10.48550/arxiv.2112.10684.
16. Ouyang, L. *et al*. Training language models to follow instructions with human feedback. *arXiv* (2022) doi:10.48550/arxiv.2203.02155.
17. ArcPrize. OpenAI o3 Breakthrough High Score on ARC-AGI-Pub. https://arcprize.org/blog/oai-o3-pub-breakthrough (2024).
18. OpenAI. OpenAI o1 Hub | OpenAI. https://openai.com/o1/ (2024).
19. DeepSeek-AI *et al*. DeepSeek-R1: Incentivizing Reasoning Capability in LLMs via Reinforcement Learning. *arXiv* (2025) doi:10.48550/arxiv.2501.12948.
20. Boiko, D. A., MacKnight, R., Kline, B. & Gomes, G. Autonomous chemical research with large language models. *Nature* **624**, 570–578 (2023).
21. Gottweis, J. *et al*. Towards an AI co-scientist. *arXiv* (2025) doi:10.48550/arxiv.2502.18864.
22. Luo, Y. *et al*. From Intention To Implementation: Automating Biomedical Research via LLMs. *arXiv* (2024) doi:10.48550/arxiv.2412.09429.
23. Su, H. *et al*. Many Heads Are Better Than One: Improved Scientific Idea Generation by A LLM-Based Multi-Agent System. *arXiv* (2024) doi:10.48550/arxiv.2410.09403.
24. Lu, C. *et al*. The AI Scientist: Towards Fully Automated Open-Ended Scientific Discovery. *arXiv* (2024) doi:10.48550/arxiv.2408.06292.
25. Swanson, K., Wu, W., Bulaong, N. L., Pak, J. E. & Zou, J. The Virtual Lab of AI agents designs new SARS-CoV-2 nanobodies. *Nature* (2025) doi:10.1038/s41586-025-09442-9.

26. Ghafarollahi, A. & Buehler, M. J. SciAgents: Automating scientific discovery through multi-agent intelligent graph reasoning. *arXiv* (2024) doi:10.48550/arxiv.2409.05556.
27. Skarlinski, M. D. *et al.* Language agents achieve superhuman synthesis of scientific knowledge. *arXiv* (2024) doi:10.48550/arxiv.2409.13740.
28. Burger, B. *et al.* A mobile robotic chemist. *Nature* **583**, 237–241 (2020).
29. Bromig, L. & Weuster-Botz, D. Accelerated adaptive laboratory evolution by automated repeated batch processes in parallelized bioreactors. *Microorganisms* **11**, (2023).
30. Singh, N. *et al.* Drug discovery and development: introduction to the general public and patient groups. *Front Drug Discov (Lausanne)* **3**, (2023).
31. Ren, F. *et al.* A small-molecule TNIK inhibitor targets fibrosis in preclinical and clinical models. *Nat. Biotechnol.* **43**, 63–75 (2025).
32. Ringel, M. S., Scannell, J. W., Baedeker, M. & Schulze, U. Breaking Eroom's Law. *Nat. Rev. Drug Discov.* **19**, 833–834 (2020).
33. Morris, M. R. *et al.* Levels of AGI: Operationalizing Progress on the Path to AGI. *arXiv* (2023) doi:10.48550/arxiv.2311.02462.
34. SAE. Taxonomy and Definitions for Terms Related to Driving Automation Systems for On-Road Motor Vehicles. https://www.sae.org/standards/content/j3016_202104 (2021).
35. Gao, S. *et al.* Empowering biomedical discovery with AI agents. *Cell* **187**, 6125–6151 (2024).
36. Tom, G. *et al.* Self-Driving Laboratories for Chemistry and Materials Science. *Chem. Rev.* **124**, 9633–9732 (2024).
37. Whittemore, R. & Melkus, G. D. Designing a research study. *Diabetes Educ.* **34**, 201–216 (2008).
38. Kumar, R. *Research Methodology: A Step-by-Step Guide for Beginners*. 440 (SAGE Publications Ltd, 2010).
39. University of Canterbury. Stages in the Research Process. https://library.canterbury.ac.nz/research-lifecycle/#.
40. Alphonse, N. The Main Stages of the Research Process - A Review of the Literature. *Int. J. Res. Rev.* **10**, 671–675 (2023).
41. Bhattacherjee, A., Toleman, M., Rowling, S., Frederiks, A. & Andersen, N. Social Science Research: Principles, Methods and Practices. *University of Southern Queensland* (2019) doi:10.26192/q7w89.
42. Schmidgall, S. *et al.* Agent Laboratory: Using LLM Agents as Research Assistants. *arXiv* (2025) doi:10.48550/arxiv.2501.04227.
43. Elbadawi, M., Li, H., Basit, A. W. & Gaisford, S. The role of artificial intelligence in generating original scientific research. *Int. J. Pharm.* **652**, 123741 (2024).
44. Davies, A. *et al.* Advancing mathematics by guiding human intuition with AI. *Nature* **600**, 70–74 (2021).
45. Liu, Z. *et al.* AIGS: Generating Science from AI-Powered Automated Falsification. *arXiv* (2024) doi:10.48550/arxiv.2411.11910.
46. Qu, Y. *et al.* CRISPR-GPT: An LLM Agent for Automated Design of Gene-Editing Experiments. *BioRxiv* (2024) doi:10.1101/2024.04.25.591003.
47. Zhou, J. *et al.* An AI Agent for Fully Automated Multi-Omic Analyses. *Adv Sci (Weinh)* **11**, e2407094 (2024).
48. Rapp, J. T., Bremer, B. J. & Romero, P. A. Self-driving laboratories to autonomously navigate the protein fitness landscape. *Nat. Chem. Eng.* **1**, 97–107 (2024).
49. Jiang, S., Evans-Yamamoto, D., Bersenev, D., Palaniappan, S. K. & Yachie-Kinoshita, A. ProtoCode: Leveraging large language models (LLMs) for automated generation of machine-readable PCR protocols from scientific publications. *SLAS Technol.* **29**, 100134 (2024).
50. Singh, G., Mishra, A., Pattanayak, C., Priyadarshini, A. & Das, R. Artificial intelligence and the Institutional Ethics Committee: A balanced insight into pros and cons, challenges, and future directions in ethical review of clinical research. *J. Integr. Med. Res.* **1**, 164 (2023).

51. Mann, S. P. *et al.* Development of Application-Specific Large Language Models to Facilitate Research Ethics Review. *arXiv* (2025) doi:10.48550/arxiv.2501.10741.
52. Sridharan, K. & Sivaramakrishnan, G. Assessing the Decision-Making Capabilities of Artificial Intelligence Platforms as Institutional Review Board Members. *J. Empir. Res. Hum. Res. Ethics* **19**, 83–91 (2024).
53. Sridharan, K. & Sivaramakrishnan, G. Investigating the capabilities of advanced large language models in generating patient instructions and patient educational material. *Eur. J. Hosp. Pharm. Sci. Pract.* (2024) doi:10.1136/ejhpharm-2024-004245.
54. Aydin, F., Yildirim, Ö. T., Aydin, A. H., Murat, B. & Basaran, C. H. Comparison of artificial intelligence-assisted informed consent obtained before coronary angiography with the conventional method: Medical competence and ethical assessment. *Digit Health* **9**, 20552076231218140 (2023).
55. Sridharan, K. & Sivaramakrishnan, G. Leveraging artificial intelligence to detect ethical concerns in medical research: a case study. *J. Med. Ethics* **51**, 126–134 (2025).
56. Szymanski, N. J. *et al.* An autonomous laboratory for the accelerated synthesis of novel materials. *Nature* **624**, 86–91 (2023).
57. King, R. D. *et al.* The automation of science. *Science* **324**, 85–89 (2009).
58. M Bran, A. *et al.* Augmenting large language models with chemistry tools. *Nat. Mach. Intell.* **6**, 525–535 (2024).
59. Huang, Y. *et al.* High-throughput microbial culturomics using automation and machine learning. *Nat. Biotechnol.* **41**, 1424–1433 (2023).
60. Wang, Y. *et al.* Automated High-Throughput Flow Cytometry for High-Content Screening in Antibody Development. *SLAS Discov.* **23**, 656–666 (2018).
61. Lu, Z. *et al.* Validation of Artificial Intelligence (AI)-Assisted Flow Cytometry Analysis for Immunological Disorders. *Diagnostics (Basel)* **14**, (2024).
62. Singh, C., Inala, J. P., Galley, M., Caruana, R. & Gao, J. Rethinking Interpretability in the Era of Large Language Models. *arXiv* (2024) doi:10.48550/arxiv.2402.01761.
63. Wang, Z. *et al.* Dataset of solution-based inorganic materials synthesis procedures extracted from the scientific literature. *Sci. Data* **9**, 231 (2022).
64. Weng, Y. *et al.* CycleResearcher: Improving Automated Research via Automated Review. *arXiv* (2024) doi:10.48550/arxiv.2411.00816.
65. Wang, Q., Xiong, Y., Zhang, Y., Zhang, J. & Zhu, Y. AutoCite: Multi-Modal Representation Fusion for Contextual Citation Generation. in *Proceedings of the 14th ACM International Conference on Web Search and Data Mining* 788–796 (ACM, 2021). doi:10.1145/3437963.3441739.
66. Lin, J., Song, J., Zhou, Z., Chen, Y. & Shi, X. Automated scholarly paper review: Concepts, technologies, and challenges. *Information Fusion* **98**, 101830 (2023).
67. Pividori, M. & Greene, C. S. A publishing infrastructure for Artificial Intelligence (AI)-assisted academic authoring. *J. Am. Med. Inform. Assoc.* **31**, 2103–2113 (2024).
68. Kuznetsov, I. *et al.* What Can Natural Language Processing Do for Peer Review? *arXiv* (2024) doi:10.48550/arxiv.2405.06563.
69. Gao, Z., Brantley, K. & Joachims, T. Reviewer2: Optimizing Review Generation Through Prompt Generation. *arXiv* (2024) doi:10.48550/arxiv.2402.10886.
70. Liu, R. & Shah, N. B. ReviewerGPT? An Exploratory Study on Using Large Language Models for Paper Reviewing. *arXiv* (2023) doi:10.48550/arxiv.2306.00622.
71. Liang, W. *et al.* Can Large Language Models Provide Useful Feedback on Research Papers? A Large-Scale Empirical Analysis. *NEJM AI* **1**, (2024).
72. Drori, I., Boston University / Columbia University, Te'eni, D. & Tel Aviv University. Human-in-the-Loop AI Reviewing: Feasibility, Opportunities, and Risks. *JAIS* **25**, 98–109 (2024).
73. BCG. Unlocking the Potential of AI in Drug Discovery. https://www.bcg.com/publications/2023/unlocking-the-potential-of-ai-in-drug-discovery (2023).

What are the limits to biomedical research acceleration through general-purpose AI?

74. Dell'Acqua, F. *et al.* Navigating the jagged technological frontier: field experimental evidence of the effects of AI on knowledge worker productivity and quality. *SSRN Journal* (2023) doi:10.2139/ssrn.4573321.
75. Noy, S. & Zhang, W. Experimental evidence on the productivity effects of generative artificial intelligence. *Science* **381**, 187–192 (2023).
76. GitHub. Research: quantifying GitHub Copilot's impact on developer productivity and happiness. https://github.blog/news-insights/research/research-quantifying-github-copilots-impact-on-developer-productivity-and-happiness/ (2022).
77. Wikipedia - The Signpost. Dispatches: Featured article writers—the inside view. https://en.wikipedia.org/wiki/Wikipedia:Wikipedia_Signpost/2008-11-24/Dispatches.
78. Rashid, D. Bulk RNA-seq Instructional Manual. https://www.montana.edu/mbi/facilities/cellular-analysis-core/additional-resources/index.html.
79. University of Mississippi Medical Center. Bioinformatics Core Services Pricing. https://umc.edu/Research/Core-Facilities/Institutional%20Core%20Facilities/Molecular-and-Genomics-Core/Services/Bioinformatics-Core-Pricing.html.
80. Kudiabor, H. Virtual lab powered by 'AI scientists' super-charges biomedical research. *Nature* (2024) doi:10.1038/d41586-024-01684-3.
81. Bao, K. *et al.* A robotic system for automated chemical synthesis of therapeutic agents. *Mater. Adv.* **5**, 5290–5297 (2024).
82. Omidvar, M. *et al.* Accelerated discovery of perovskite solid solutions through automated materials synthesis and characterization. *Nat. Commun.* **15**, 6554 (2024).
83. Delgado-Licona, F. & Abolhasani, M. Research Acceleration in Self-Driving Labs: Technological Roadmap toward Accelerated Materials and Molecular Discovery. *Advanced Intelligent Systems* 2200331 (2022) doi:10.1002/aisy.202200331.
84. Emerald cloud lab. Publish Faster & Cheaper with a Cloud Lab. https://www.emeraldcloudlab.com/why-cloud-labs/efficiency/academia/ (2025).
85. Arnold, C. Cloud labs: where robots do the research. *Nature* **606**, 612–613 (2022).
86. Martin, H. G. *et al.* Perspectives for self-driving labs in synthetic biology. *Curr. Opin. Biotechnol.* **79**, 102881 (2023).
87. Volk, A. A. & Abolhasani, M. Performance metrics to unleash the power of self-driving labs in chemistry and materials science. *Nat. Commun.* **15**, 1378 (2024).
88. Arias, D. S. & Taylor, R. E. Scientific discovery at the press of a button: navigating emerging cloud laboratory technology. *Adv. Mater. Technol.* **9**, (2024).
89. Brooks, S. M. & Alper, H. S. Applications, challenges, and needs for employing synthetic biology beyond the lab. *Nat. Commun.* **12**, 1390 (2021).
90. Seifrid, M. *et al.* Autonomous Chemical Experiments: Challenges and Perspectives on Establishing a Self-Driving Lab. *Acc. Chem. Res.* **55**, 2454–2466 (2022).
91. Qian, L., Dong, Z. & Guo, T. Grow AI virtual cells: three data pillars and closed-loop learning. *Cell Res.* (2025) doi:10.1038/s41422-025-01101-y.
92. Timmers, M. *et al.* How do 66 European institutional review boards approve one protocol for an international prospective observational study on traumatic brain injury? Experiences from the CENTER-TBI study. *BMC Med. Ethics* **21**, 36 (2020).
93. Manganaro, L. The true latency of biomedical research papers. *Scientometrics* **129**, 2897–2910 (2024).
94. Andersen, M. Z., Fonnes, S. & Rosenberg, J. Time from submission to publication varied widely for biomedical journals: a systematic review. *Curr. Med. Res. Opin.* **37**, 985–993 (2021).
95. Huisman, J. & Smits, J. Duration and quality of the peer review process: the author's perspective. *Scientometrics* **113**, 633–650 (2017).
96. Aczel, B., Szaszi, B. & Holcombe, A. O. A billion-dollar donation: estimating the cost of researchers' time spent on peer review. *Res. Integr. Peer Rev.* **6**, 14 (2021).

# Supplementary information

## Supplementary Table S1. GPAI capabilities in research tasks

| Major research task | Exemplary sub-tasks | Exemplary work |
|---|---|---|
| Knowledge synthesis | Finding & curating | (Skarlinski et al. 2024) Synthesizes scientific knowledge by retrieving relevant papers and summarizing their content in a cited, Wikipedia-style format. Outperforms human experts in precision and provides a more structured and accurate synthesis of scientific literature.<br><br>(Ghafarollahi and Buehler 2024) Knowledge graphs + agents can create novel hypotheses for science and rank them for novelty and feasibility.<br><br>(Z. Wang et al. 2022) Automated extraction of synthesis protocols from scientific literature to combat information overload through intelligent filtering and aggregation of research data.<br><br>(Luo et al. 2024) Automated literature search across multiple databases, standardization of research papers into experimental reports, and analysis of literature relevance and usability.<br><br>(Schmidgall et al. 2025) The PhD-agent retrieves relevant literature using the arXiv API. It uses iterative querying, evaluating abstracts, and full-text analysis to curate a set of high-quality research papers relevant to the research idea. |
| | Critical evaluation | (Elbadawi et al. 2024) Demonstrates critical thinking, predicting effects like laser scanning speed on printlet properties without prior precedent or templates in the literature.<br><br>(Swanson et al. 2025) Conducts critical evaluation through virtual team meetings involving multiple agents, including a dedicated Critic agent.<br><br>(Schmidgall et al. 2025) Assesses papers during literature reviews for relevance and importance, aligning findings with research goals and enabling automated summarization and cross-referencing for experimental planning.<br><br>(Ghafarollahi and Buehler 2024) Employs a Critic agent to review research proposals, highlighting strengths, weaknesses, and areas for improvement. |
| | Synthesize findings | (Swanson et al. 2025) Employs multiple AI agents collaboratively, led by a Principal Investigator agent overseeing project coordination.<br><br>(Schmidgall et al. 2025) Integrates curated literature into frameworks for experimental design and hypothesis development via human-agent collaboration or autonomous processing.<br><br>(Ghafarollahi and Buehler 2024) Expands hypotheses systematically, synthesizing findings into structured, comprehensive research outputs. |
| | Gap & contradiction identification | (Skarlinski et al. 2024) Detects contradictions in scientific literature by extracting claims, comparing them using a contradiction-detection prompt, scoring on a Likert scale, and validating with expert review.<br><br>(Ghafarollahi and Buehler 2024) Utilizes multi-agent systems and knowledge graphs to autonomously generate, critique, and refine hypotheses, identifying contradictions and gaps through structured data analysis and novelty assessment tools like the Semantic Scholar API. |
| Idea & hypothesis generation | Problem identification | (Ren et al. 2025) Generated hypotheses through text mining and multiple data sources, calculated success probabilities, identified TNIK as optimal target and progressed to Phase 2 trials.<br><br>(Schmidgall et al. 2025) Human researchers input a broad research problem, which the system refines into actionable goals. The PhD and Postdoc agents expand on this idea to create a structured research question, identifying specific opportunities for testing.<br><br>(Davies et al. 2021) Machine learning guides mathematical intuition and aids in faster discovery of new conjectures and theorems.<br><br>(H. Su et al. 2024) Multi-agent collaboration improves hypothesis generation quality, while reducing computing costs.<br><br>(Z. Liu et al. 2024) Specialized Agents produce testable hypotheses.<br><br>(C. Lu et al. 2024) The AI Scientist generates hypotheses, scores them for novelty and feasibility, and refines ideas using iterative chain-of-thought and self-reflection mechanisms. |
| | Hypothesis formulation | (Skarlinski et al. 2024) By flagging contradictions or evidence gaps these LLM-based literature agents can support the generation of new hypotheses.<br><br>(Ghafarollahi and Buehler 2024) AI uses graph reasoning to identify gaps, propose solutions, and ensure rigor and falsifiability through critique agents.<br><br>(Z. Liu et al. 2024) A 'falsification agent' verifies or refutes scientific claims by designing and executing automated ablation studies.<br><br>(C. Lu et al. 2024) Hypotheses are shaped into experimentally testable goals, ensuring alignment with research objectives. |
| | Theoretical framework development | (Ren et al. 2025) PandaOmics mapped pathways for lung fibrosis and cancer hallmarks, synthesizing literature into actionable solutions.<br><br>(Ghafarollahi and Buehler 2024) AI synthesizes knowledge graphs, links constructs, and provides mechanistic explanations, similar to PandaOmics.<br><br>(Z. Liu et al. 2024) Works within and refines theoretical frameworks through ablation testing and falsification to establish ground truths.<br><br>(Schmidgall et al. 2025) The agents build a framework around the research hypotheses, linking them to experimental design and objectives.<br><br>(Swanson et al. 2025) Inter-disciplinary agent meetings iteratively refine the project plan and tool chain. |
| | Feasibility assessment | (Ren et al. 2025) PandaOmics identified TNIK as the best target by analyzing pathways and calculating causal inferences.<br><br>(Ghafarollahi and Buehler 2024) AI evaluates hypotheses for novelty, practicality, and alignment with literature while proposing validation strategies.<br><br>(Swanson et al. 2025) Critique agents reviews practicality of code and methodological choices.<br><br>(Z. Liu et al. 2024) Experimental executability is ensured by translating proposals into structured, errorless instructions.<br><br>(Schmidgall et al. 2025) Automated troubleshooting and iterative refinement during experimentation ensure that research plans are executable.<br><br>(C. Lu et al. 2024): Assesses hypotheses for novelty and practicality, dynamically iterating on experiments to ensure executable research plans. |
| **Experiment design** | Method selection | (Ghafarollahi and Buehler 2024) AI agents systematically recommend tools and develop precise experimental and synthesis protocols.<br><br>(Swanson et al. 2025) Reasons across multiple non related disciplines (Biology, CS) to automatically design functional nanobodies using a computational pipeline (ESM, AlphaFold-Multimer, Rosetta).<br><br>(Schmidgall et al. 2025) The system formulates a detailed experimental plan, defining variables, objectives, methods and expected results. Plans are informed by literature review outputs and aligned with hypotheses to ensure reliable experiments.<br><br>(Rapp, Bremer, and Romero 2024) The AI autonomously designs experiments by predicting optimal protein sequences using Bayesian optimization, selecting candidates based on model predictions, and specifying experimental protocols tailored to test these hypotheses.<br><br>(Qu et al. 2024) CRISPR-GPT automates experiment setup by selecting CRISPR systems, designing guide RNAs, and tailoring delivery methods based on user objectives.<br><br>(Luo et al. 2024) The AI uses RAG and hierarchical learning to analyze literature and datasets, identifying experimental frameworks and variables.<br><br>(C. Lu et al. 2024) The AI generates experiment designs based on initial templates, existing literature, and self-generated hypotheses. It plans experiments iteratively, incorporating feedback from results to improve designs.<br><br>(Zhou et al. 2024) AutoBA generates detailed, customized analysis plans by leveraging user-provided data paths, descriptions, and objectives. |
| | Protocol development | (Ghafarollahi and Buehler 2024) Proposes detailed experimental protocols in defined steps.<br><br>(Swanson et al. 2025) Develops protocols for automated ESM mutation analysis, AlphaFold structure prediction, and Rosetta calculations.<br><br>(Jiang et al. 2024) ProtoCode automates the curation and standardization of protocols from unstructured text.<br><br>(Schmidgall et al. 2025) AI designs protocols to run experiments autonomously, ensuring clarity and reproducibility.<br><br>(Qu et al. 2024) Generates detailed protocols, including gRNA design, delivery setups, and off-target prediction, while integrating external tools for resource optimization.<br><br>(Luo et al. 2024) The AI designs detailed protocols step-by-step, including headings, outlines, and experimental details based on analyzed data. |
| | Quality control | (Jiang et al. 2024) Standardizes protocols to ensure quality.<br><br>(Schmidgall et al. 2025) Automated error handling and iterative self-reflection ensure robust experimentation. Quality checks are integrated into every stage, including debugging and runtime error detection during data preparation.<br><br>(Ghafarollahi and Buehler 2024) Iterative feedback ensures protocols are robust and aligned with the hypotheses, including clear steps for modeling, synthesis, and testing.<br><br>(Qu et al. 2024) CRISPR-GPT provides validation workflows, such as sequencing, functional assays, and off-target analysis, to ensure experimental accuracy and compliance.<br><br>(Luo et al. 2024) The AI employs an LLM-based reviewer to ensure protocols meet quality metrics like completeness, correctness, and logical soundness. |
| Ethics approval & permits | Initial screening | (G. Singh et al. 2023) AI can enhance efficiency and standardization by rapidly analyzing documents and applying consistent criteria to identify potential ethical issues.<br>(Sridharan and Sivaramakrishnan 2025) LLMs can streamline ethics review processes by helping institutional review board members evaluate protocols more efficiently.<br>(Sridharan and Sivaramakrishnan 2024a) AI can speed up initial screening by quickly identifying good clinical practice violations and standard operating procedure deficiencies.<br>(Mann et al. 2025) An LLM can automatically review submissions for completion and ethics issues, suggest categories, and highlight where additional details are needed. |
| | Scientific review | (G. Singh et al. 2023) AI algorithms can provide a comprehensive perspective by exploring and cross-referencing databases of research studies and ethical guidelines.<br>(Sridharan and Sivaramakrishnan 2025) LLMs can systematically evaluate research proposals against guidelines and flag compliance issues, while complex ethical decisions require human expertise.<br>(Sridharan and Sivaramakrishnan 2024a) AI can enhance scientific rigor by assessing study design and eligibility for expedited review.<br>(Mann et al. 2025) LLMs can assist in scientific review by summarizing study aims, identifying key design elements, and highlighting ethical considerations. |
| | Ethics assessment | (Aydin et al. 2023) Physician-delivered informed consent was compared to an AI-based approach, which achieved better patient understanding while maintaining satisfaction levels.<br>(G. Singh et al. 2023) AI can assist in risk assessment and the review of the informed consent process, but requires human oversight to address complex issues.<br>(Sridharan and Sivaramakrishnan 2025) LLMs demonstrated ability to identify ethical issues in case studies but performed suboptimally in assessing placebo use, risk mitigation, and participant risks.<br>(Sridharan and Sivaramakrishnan 2024a) AI can provide a more consistent ethics review, but human oversight remains crucial.<br>(Mann et al. 2025) An LLM could provide a preliminary review, identifying ethical issues, precedents and guidelines, and a risk-benefit assessment. |
| | Regulatory compliance | (G. Singh et al. 2023) AI could facilitate transnational collaboration by supporting adherence to guidelines and regulations, using common tools and criteria across jurisdictions.<br>(Sridharan and Sivaramakrishnan 2025) AI can assist with adherence to guidelines by identifying missing elements and assessing fundamental ethical issues in research proposals.<br>(Sridharan and Sivaramakrishnan 2024a) AI can facilitate adherence to standards by drafting standard operating procedures, but human adaptation is needed.<br>(Mann et al. 2025) LLMs could support regulatory compliance by comparing protocols with applicable regulations, and institutional policies, also considering national or local context. |
| | Administrative processing and monitoring | (Aydin et al. 2023) AI streamlines the process of gathering and presenting information to patients, reducing clinical workload and costs.<br>(G. Singh et al. 2023) AI can automate administrative tasks, maintain documentation, and support continuous learning and adaptation to new findings.<br>(Sridharan and Sivaramakrishnan 2025) LLMs offer cost-effective solutions by generating initial drafts and training materials but still require human review and editing.<br>(Sridharan and Sivaramakrishnan 2024a) AI offers cost-effective solutions by streamlining standard operating procedure creation and assisting in the administrative tasks, requiring human validation.<br>(Sridharan and Sivaramakrishnan 2024b) LLMs show potential to automate patient instructions and materials, though successful implementation requires proper medical oversight.<br>(Mann et al. 2025) LLMs could perform consistency checks against past decisions, institutional policies, and also flag inconsistencies in documentation. |
| **Experiment execution** | Experiment preparation | (Szymanski et al. 2023) Robots automate the preparation of materials, precise measurement, and transfer to ensure experiments start with minimal human intervention.<br>(King et al. 2009) The system autonomously designs experiments by selecting yeast strains and preparing growth mediums based on hypotheses.<br>(Rapp, Bremer, and Romero 2024) The SAMPLE platform integrates automated workflows for assembling DNA, preparing reagents, and configuring robotic systems for protein engineering experiments.<br>(Bromig and Weuster-Botz 2023) AI handles the transfer of media and inoculum for serial passaging between bioreactors, ensuring precision and reducing manual effort.<br>(M Bran et al. 2024) Leverages LLMs to integrate chemistry tools for chemical discovery, synthesis planning, and reaction prediction.<br>(Huang et al. 2023) AI automates phenotypic data collection and colony isolation, reducing manual setup and standardizing processes.<br>(Y. Wang et al. 2018) Modular robotic systems integrate precise timing and automated instruments to streamline sample handling, ensure scalability, and deliver reproducible execution of protocols. |
| | Experiment execution | (Burger et al. 2020) Executes fully automated multi-step workflows, managing parallel setups of 16 samples with precision, achieving high acceleration in experimental processes.<br>(Jiang et al. 2024) Converts protocols into machine-readable formats for lab equipment.<br>(Szymanski et al. 2023) Robots in the A-Lab autonomously execute tasks such as synthesis, heating, cooling, and data collection, integrating physical actions with AI-driven decision-making to ensure precision, consistency, and minimal human intervention in experimental workflows.<br>(King et al. 2009) It uses laboratory automation to physically execute the experimental plan, including inoculating strains, managing growth conditions, and monitoring growth curves.<br>(Rapp, Bremer, and Romero 2024) The platform autonomously executes reproducible protocols like PCR amplification, thermostability assays, gene assembly, protein expression, and biochemical evaluations.<br>(Bromig and Weuster-Botz 2023) Real-time monitoring and automated consistent experiment execution accelerate processes, reducing time compared to manual methods.<br>(Huang et al. 2023) Robotic systems accelerate colony picking, imaging, and genotyping, improving experimental throughput.<br>(Y. Wang et al. 2018) The iQue PLUS Screener enables high-throughput data acquisition, reducing experiment runtime. |
| | Experiment documentation | (Luo et al. 2024) The AI retrieves, filters, and processes datasets and literature into structured reports that serve as input for protocol design.<br>(Burger et al. 2020) The system collects process data and can be monitored remotely, with data stored for analysis.<br>(Jiang et al. 2024) Standardizes how protocols are recorded.<br>(Szymanski et al. 2023) Implemented automated collection and storage of XRD patterns, synthesis conditions, and reaction outcomes through an integrated control system. |
| | Troubleshooting and optimization | (Burger et al. 2020) Incorporates 24/7 CCTV monitoring, remote error resolution, automatic alerts for stock levels and failures, and Bayesian optimization for outcome improvement.<br>(Szymanski et al. 2023) Created an active learning system that could optimize failed syntheses by suggesting improved reaction pathways, though still requiring human oversight for some failure modes.<br>(King et al. 2009) Adam's ability to cycle through hypothesis testing and refine its approach highlights its optimization capabilities. |
| | | (Rapp, Bremer, and Romero 2024) The system iteratively refines its protein engineering process through Bayesian optimization, improving the experimental design based on feedback from previous results. |
| | | (Bromig and Weuster-Botz 2023) AI uses real-time data and soft sensors to optimize growth conditions and reduce lag phases. |
| | | (Huang et al. 2023) Morphological analysis ensures accurate colony identification, enhancing data quality and reproducibility. |
| | | (Y. Wang et al. 2018) Iterative system refinements and flexible automation configurations minimize cell loss and maximize workflow efficiency. |
| | Material management | (Burger et al. 2020) Manages the sample lifecycle, including preparation, analysis, and storage, using an organized rack system to maintain sample integrity. |
| | | (King et al. 2009) The system utilizes an automated freezer and manages sample storage as part of its workflow. |
| | | (Szymanski et al. 2023) Robotics enable secure, efficient sample handling and storage throughout the experimental workflow. |
| | Equipment maintenance | (Burger et al. 2020) Handles positioning calibration and battery management with automated systems. |
| | | (Huang et al. 2023) The system integrates biobanking with searchable databases, enabling efficient data storage and retrieval. |
| | | (Y. Wang et al. 2018) Automated systems use a modular robotic platform and integrated instruments to reduce manual intervention in sample preparation. |
| Data analysis | Data collection | (Schmidgall et al. 2025) Data preparation is automated using the ML Engineer agent, which writes and validates data processing scripts. |
| | | (Ghafarollahi and Buehler 2024) The AI system proposes detailed protocols for experiments and molecular simulations, including specifying the types of data to be collected, such as binding energies and self-assembly structures. |
| | | (Qu et al. 2024) The system designs validation protocols, including methods selection and the design of primers, to guide the collection of experimental outcomes. |
| | | (C. Lu et al. 2024) The AI autonomously gathers, processes, and organizes experimental results, generating visualizations and structured summaries for scientific use. |
| | Data cleaning and organisation | (Zhou et al. 2024) Automates preprocessing steps, including quality control, adapter trimming, and alignment. |
| | | (Z. Lu et al. 2024) DeepFlow performs an integrity check on input LMD files, automatically removes debris and doublets, and standardizes datasets for analysis, replacing manual preprocessing steps with automated workflows. |
| | | (Schmidgall et al. 2025) Data cleaning is automated during the preparation phase to ensure high-quality inputs for experiments. The system uses Python scripts to format and preprocess datasets. |
| | | (Qu et al. 2024) The system manages and validates user inputs, particularly when handling guide RNA designs and sequencing information. |
| | | (Swanson et al. 2025) The Virtual Lab AI agents creates an iterative loop of selection and improvement of most promising candidates. |
| | Statistical analysis and pattern identification | (Zhou et al. 2024) AutoBA eliminates traditional bioinformatics complexity by autonomously handling the entire analysis pipeline - from tool selection to code execution - requiring only the input data and desired analysis goal. |
| | | (Luo et al. 2024) Utilizes an LLM-based agent to design and execute dry lab experiments, automating bioinformatics analyses based on experimental protocols. |
| | | (Schmidgall et al. 2025) Employs an mle-solver to detect meaningful patterns and validate experimental results against clearly defined scientific objectives, ensuring robust outcomes. |
| | | (Z. Lu et al. 2024) Using clustering algorithms, the AI efficiently identifies and classifies normal and abnormal cell populations. |
| | Data visualization | (Z. Lu et al. 2024) The AI generates clear, multidimensional scatterplots for comprehensive and interpretable results. |
| | | (Swanson et al. 2025) AI agents used tools like AlphaFold-Multimer, Rosetta, and ESM that could later be used to produce high quality visualizations. |
| | | (C. Lu et al. 2024) The AI Scientist generates figures and plots as part of its experimental workflow using Python scripts. |
| | | (C. Singh et al., 2024) AI powers data analysis by transforming complex datasets into clear explanations and insightful visualizations, enabling reliable interpretation and extraction of key findings. |
| | | (Schmidgall et al. 2025) The AI models generated two figures within the set limit. |
| | Validation analysis | (Zhou et al. 2024) Ensures reliability of code execution and results through an automated error-checking mechanism (e.g., automatic code repair). |
| | | (Schmidgall et al. 2025) Outputs are validated by comparing experimental results to hypotheses and scoring their alignment. Scoring functions assess the quality and effectiveness of results. |
| | | (Swanson et al. 2025) The Virtual Lab AI agents employ bioinformatic software modules to score nanobody candidates. |
| Results interpretation | Results synthesis | (Ghafarollahi and Buehler 2024) The system employs multi-agent AI to critically analyze and interpret data from a knowledge graph, providing detailed outcomes and mechanisms for proposed hypotheses. |
| | Hypothesis evaluation | (Schmidgall et al. 2025) The agents discuss and evaluate whether the experimental results support or refute the hypotheses. |
| | | (Ghafarollahi and Buehler 2024) The system includes Critic agents that evaluate hypotheses against novelty and feasibility using tools like the Semantic Scholar API. |
| | Result contextualization | (Schmidgall et al. 2025) Agents outline the broader implications of the findings in the discussion section of the report, guiding future research directions. |
| | | (Ghafarollahi and Buehler 2024) The AI demonstrates contextual awareness and adaptability in hypothesis generation by dynamically integrating agent interactions, incorporating human feedback for refinement, and leveraging tools like the Semantic Scholar API to ensure novelty and relevance in scientific ideas. |
| **Manuscript preparation** | Methods documentation | (Elbadawi et al. 2024) The AI was able to write a full methodology section, with detailed experimental protocols. <br> (C. Lu et al. 2024) The AI Scientist documents the content of each plot, ensuring the saved figures and experimental notes contain all the necessary information for drafting the paper. <br> (Weng et al. 2024) AI generates logically coherent, domain-specific texts and simulated experimental protocols step-by-step. <br> (Zhou et al. 2024) AutoBA is a transparent and interpretable tool that allows bioinformaticians to easily document, modify, and customize its methods, streamlining the data analysis process. |
| | Results presentation | (Elbadawi et al. 2024) The AI effectively authored the results section, embedding scientific insights and methodologies, including experimental protocols. <br> (Schmidgall et al. 2025) The paper-solver generates an initial scaffold, dividing the manuscript into standard academic sections. <br> (C. Lu et al. 2024) The AI scientist motivates, explains, and summarizes results with complete visualizations. |
| | Reference management | (Skarlinski et al. 2024) The system enforces inline citations by requiring citation identifiers for each assertion and limiting citations to the provided context, thereby improving the completeness and accuracy of citations. <br> (C. Lu et al. 2024) The AI Scientist automatically searches for relevant papers using the Semantic Scholar API, integrates citations during the write-up, and generates a complete reference list. <br> (Weng et al. 2024) Outperforms the AI Scientist by citing more papers, enabling a deeper understanding of related work. <br> (Q. Wang et al. 2021) AutoCite demonstrates that reliable citation and context generation in academic papers is achievable by integrating semantic and structural insights. |
| | Figure and table preparation | (Elbadawi et al. 2024) The AI created believable and compelling analytical data, including plots and photo-realistic images of the subject matter. <br> (Schmidgall et al. 2025) The AI models successfully generated two figures, though their output was constrained to a maximum of two. <br> (C. Lu et al. 2024) The AI Scientist creates figures by leveraging automated tools such as Aider, an advanced LLM-based coding assistant, which edits plotting scripts and generates visualizations based on experimental results. |
| | Data sharing | (Schmidgall et al. 2025) Creates sharable code repository for reproducibility. <br> (C. Lu et al. 2024) Saves experiment results and logs in reproducible formats for collaboration and transparency. |
| | Writing and revision | (Skarlinski et al. 2024) Generates Wikipedia-style articles with cited summaries. While not explicitly described as manuscript preparation, this is closely related to creating structured, referenced scientific content. <br> (Elbadawi et al. 2024) Was able to write a full publication-ready manuscript on GPT-4 based on its own synthetically generated data, including tables and figures. <br> (Weng et al. 2024) CycleResearcher can generate fully structured research papers with clear methodological descriptions, deliver reviews with detailed feedback across multiple criteria, and consistently evaluate and refine work through iterative feedback. <br> (C. Lu et al. 2024) Refines manuscripts iteratively using self-assessment and simulated peer-review feedback. <br> (Schmidgall et al. 2025) The system produces complete, submission-ready academic reports adhering to NeurIPS formatting standards, refining manuscripts through iterative edits for clarity and coherence, with LaTeX compilation ensuring document integrity. |

| Publication process | Journal selection & submission | (Pividori and Greene 2024) AI-assisted writing tools can reduce the burden of formatting and stylistic requirements in scientific writing. (Lin et al. 2023) AI could help with journal selection by automated scope-evaluation, potentially reducing desk rejections. (Kuznetsov et al. 2024) AI could help generate publication metadata like keywords and track suggestions, and reformat submissions for different presentation styles. |
|---|---|---|
| | Screening & reviewer assignment | (Lin et al. 2023) AI tools can speed up manuscript screening by evaluating submissions for format, plagiarism, and article type. (R. Liu and Shah 2023) LLMs like GPT-4 can be used to verify author-provided checklists with a demonstrated high accuracy covering topics such as theoretical results, experimental results, and code. (Kuznetsov et al. 2024) Automated screening systems can evaluate manuscripts for formatting and policy compliance while optimizing reviewer assignments through improved keyword and content similarity matching. |
| | Peer review | (Lin et al. 2023) While AI-assisted systems will initially complement human reviewers, automated scholarly paper review could ultimately manage the entire evaluation process once challenges in data, parsing, interaction, and reasoning are resolved. (R. Liu and Shah 2023) LLMs like GPT-4 have demonstrated error detection capabilities in scientific papers, identifying both mathematical and conceptual errors in test papers. (Liang et al. 2024) AI reviewers provide paper-specific feedback, showing similar overlap with human reviews as found between human reviewers. (Gao, Brantley, and Joachims 2024) LLM-generated reviews can be enhanced using aspect prompts to focus on specific parts of a paper, leading to detailed feedback that covers a range of opinions. (Drori et al. 2024) Human judges rated LLM-generated reviews as comparable in quality to human reviews, though LLMs cannot yet handle all review cases independently. (Weng et al. 2024) CycleReviewer demonstrates expert-level review capabilities, providing detailed, consistent feedback across multiple criteria and integrating iterative feedback throughout the research-review-revision cycle. (C. Lu et al. 2024) The AI Scientist generates an automated review, according to current practice at standard machine learning conferences. |
| | Revision | (Pividori and Greene 2024) While the AI-based Manubot Editor successfully enhanced most paragraphs, some revisions introduced errors or omitted key information. (Lin et al. 2023) Automated scholarly paper review can offer immediate feedback, enabling faster and more efficient revisions and improvements to manuscripts. (Liang et al. 2024) LLMs can provide constructive feedback and suggestions for enhancing manuscripts, more than half of the users found the LLM generated feedback helpful. (Drori et al. 2024) LLMs receive ratings equivalent to human reviewers in how helpful their reviews are at guiding authors toward paper improvements. |

## Supplementary Table S2. Acceleration factors overview

| Acceleration Range | Acceleration (Arithmetic Mean) | Task Type | Reference |
|---|---|---|---|
| 1.3-2.0 | 1.65 | Cognitive | BCG |
| 2.0-4.0 | 3 | Cognitive | Lu et al. |
| 1.3 | 1.3 | Cognitive | Dell'Acqua et al. |
| 1.7 | 1.7 | Cognitive | Noy & Zhang |
| 2.2 | 2.2 | Cognitive | GitHub Copilot |
| 75-300 | 187.5 | Cognitive | Skarlinski et al. |
| 120-140 | 130 | Cognitive | Zhou et al. |
| 150-300 | 225 | Cognitive | Luo et al. |
| 1.7 | 1.7 | Physical | Bao et al. |
| 1-2 | 1.5 | Physical | Rapp et al. |
| 3-6 | 4.5 | Physical | Rapp et al. |
| 15-50 | 32.5 | Physical | Rapp et al. |
| 20 | 20 | Physical | Huang et al. |
| 12-36 | 24 | Physical | Omidvar et al. |
| 40 | 40 | Physical | Burger et al. |
| 10-100 | 55 | Physical | Burger et al. |
| 10-100 | 55 | Physical | Delgado-Licona & Abolhasani |
| 2 | 2 | Physical | Emerald cloud lab |
| 90 | 90 | Physical | Emerald cloud lab |
| 100 | 100 | Physical | Arnold |

## Supplementary Table S3. Expert survey: Project timelines

| Total project | Knowledge synthesis | Idea & hypothesis generation | Experiment design | Ethics approval & permits | Experiment execution | Data analysis | Results inter-pretation | Manuscript preparation | Publication process |
|---|---|---|---|---|---|---|---|---|---|
| 87 | 5 | 5 | 5 | 0 | 30 | 19 | 5 | 12 | 6 |
| 72 | 2 | 2 | 2 | 2 | 24 | 8 | 8 | 6 | 18 |
| 80 | 2 | 3 | 5 | 3 | 19 | 21 | 2 | 5 | 20 |
| 80 | 3 | 8 | 6 | 6 | 18 | 9 | 6 | 6 | 18 |
| 66 | 2 | 2 | 12 | 0 | 18 | 6 | 6 | 8 | 12 |
| 118 | 6 | 6 | 4 | 4 | 36 | 14 | 14 | 6 | 28 |
| 40 | 1 | 1 | 1 | 2 | 8 | 12 | 5 | 4 | 6 |
| 36 | 2 | 3 | 2 | 1 | 6 | 10 | 1 | 10 | 1 |

## Supplementary Table S4. Expert survey: Acceleration plausibility

| Knowledge synthesis | Idea & hypothesis generation | Experiment design | Ethics approval & permits | Experiment execution | Data analysis | Results interpretation | Manuscript preparation | Publication process |
|---|---|---|---|---|---|---|---|---|
| Moderate Overestimate | Significant Overestimate | Moderate Overestimate | Moderate Overestimate | Moderate Overestimate | Moderate Overestimate | Plausible Estimate | Moderate Overestimate | Moderate Overestimate |
| Plausible Estimate | Moderate Overestimate | Significant Overestimate | Moderate Underestimate | Moderate Overestimate | Plausible Estimate | Moderate Overestimate | Plausible Estimate | Plausible Estimate |
| Moderate Overestimate | Moderate Overestimate | Moderate Overestimate | Plausible Estimate | Significant Overestimate | Plausible Estimate | Plausible Estimate | Plausible Estimate | Moderate Overestimate |
| Moderate Overestimate | Moderate Overestimate | Significant Overestimate | Plausible Estimate | Moderate Overestimate | Plausible Estimate | Moderate Overestimate | Plausible Estimate | Plausible Estimate |
| Plausible Estimate | Moderate Overestimate | Moderate Overestimate | Moderate Underestimate | Moderate Overestimate | Moderate Overestimate | Moderate Overestimate | Plausible Estimate | Plausible Estimate |
| Significant Overestimate | Plausible Estimate |  | Plausible Estimate | Moderate Overestimate | Plausible Estimate | Moderate Overestimate | Moderate Underestimate | Moderate Underestimate |
| Significant Overestimate | Moderate Overestimate | Moderate Overestimate | Plausible Estimate | Significant Overestimate | Moderate Overestimate | Moderate Overestimate | Moderate Overestimate | Moderate Overestimate |
| Plausible Estimate | Moderate Overestimate | Significant Overestimate |  | Plausible Estimate | Plausible Estimate | Moderate Underestimate | Significant Underestimate | Significant Underestimate |

## Supplementary Table S5. Expert survey: Limiting factors

| Biological/ Physical time limits | Resource & infrastructure | Input data limitations | Human strategic direction | Human ethical judgment | Human accountability | Institutional adaptation | Empirical validation | Stakeholder coordination | Safety & security | Scientific community assimilation | Data volume management |
|---|---|---|---|---|---|---|---|---|---|---|---|
| Moderate | Major | Crucial | Major | Major | Crucial | Crucial |  | Moderate | Crucial | Major | Crucial |
| Crucial | Moderate | Insignificant | Insignificant | Insignificant | Insignificant | Minor | Minor | Moderate | Insignificant | Major | Moderate |
| Crucial | Minor | Moderate | Major | Major | Crucial | Major | Moderate | Crucial | Minor | Crucial | Major |
| Insignificant | Crucial | Minor | Major | Major | Crucial | Minor | Moderate | Minor | Major | Major | Insignificant |
| Minor | Minor | Moderate | Minor | Major | Minor | Moderate | Crucial | Moderate | Moderate | Major | Insignificant |
| Crucial |  | Major | Minor | Minor | Moderate | Major | Moderate | Major | Moderate | Moderate | Minor |
| Major | Moderate | Crucial | Minor | Moderate | Minor | Moderate | Major | Moderate | Moderate | Moderate | Minor |
| Moderate | Moderate | Crucial | Minor | Major | Major | Crucial | Major | Moderate | Crucial | Crucial | Major |

# Supplementary Table S6. Expert survey: Open comments

## Other limiting factors

*Response 1:* Interfacing of various output/input systems - ideally no human in chain - likely requires a single system handling AI and robotics or very limited, highly standardised systems? Maybe I am thinking too non-AI like and a human-like, general intelligence operating available individual robot systems can immediately use a non-roboter lab better than a human. Cost/Benefit ratio - upfront cost to set up such a system requires phenomenal trust in results (i.e. monetization (or "academic currency")) or one ends up with a much weaker patchwork of 2nd choice systems that have downsides, don't interact etc with each other. Though similar to e.g. black powder's effect in warfare, I assume once it is established, it is pretty much needed to keep up/stay relevant. *(Rated as: Moderate limit)*
- - - - -

*Response 2:* Fundamental Biological/Physical/Chemical Time & Measurement Limits: In my research, the blood sampling of 200 individuals simply takes a definite time, which can not be shortened due to ethical reasons and since blood drawing cannot be faster even with maximum AI. *(Rated as: Crucial limit)*
- - - - -

*Response 3:* Fundamental time-frame of the experiment (i.e. looking at 3 month effect after intervention) and schedules of the research subjects. *(Rated as: Crucial limit)*

## General considerations

*Response 1:* "revisions" was part of two categories Some of the categories have common parts (e.g. revisions could also be a part of "data analysis" and "results interpretation")
- - - - -

*Response 2:* While AI might accelerate data analyses and presentation, the speed of publication with peer review and also the response of the co-authors cannot be changed….
- - - - -

*Response 3:* The project I used as reference for making the time estimates was riddled with errors in project management, e.g. 12 months of experimental execution were spent on trying to optimize the wrong thing - similar to hitchhikers guide through the galaxy we might ask the wrong questions/invest in the wrong avenues. I am not sure if human-level AI (faster but same "intellect") would be able to help with that? Though if early adopters (of a robotics lab headed by AI) do this, the potential to "waste" money faster than a human lab (i.e. by running 24/7) is high. Human Accountability & Responsibility I currently answered as a crucial limit, but this is a societal decision, can we relent control? (Humans likely won't want to be responsible for black-box processes - i.e. be in a scapegoat position). I guess a significant benefit comes from running 24/7, being able to utilize logistics efficiently (machines constantly running, many projects in parallel). I.e. emerging efficiencies of "scale" by one/many general purpose AIs coordinating instead of humans. Regarding time savings, perhaps a usable estimator is the difference between an "old Mom-and-Pa bookshop" and Amazon? On the experimental side there is A LOT of logistical inefficiency in most labs. (I have not checked the provided references, I assume they consider these aspects.) Overall, I would not give a lot of weight to my predictions/estimations and expect to be wrong about most of them.

# Supplementary Information S7. Expert Survey: Interface (PDF Archive)

Expert Elicitation: Limits to accelerating research through future AI

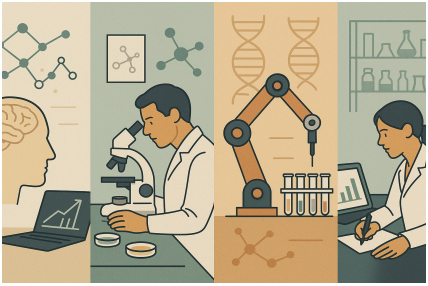

**Introduction**

**"General-purpose AI"** refers to AI systems that are capable of competently performing a wide range of distinct tasks. Rapid advances in such general-purpose AI, including powerful language models and sophisticated AI agents integrated with robotics and automated "cloud labs", are beginning to **reshape biomedical research**. Evidence suggests these technologies hold immense potential to accelerate scientific discovery.

**General-purpose AI systems are demonstrating capabilities across the research lifecycle:**

- **Cognitive tasks:** AI agents can rapidly synthesize vast amounts of literature, generate novel hypotheses, design complex experiments, analyze large datasets, and even draft manuscripts, sometimes achieving significant speedups (e.g., Skarlinski 2024a reported **100x** faster knowledge synthesis; Luo 2024a reported **150-300x** faster research cycles in specific contexts)**.**

- **Physical tasks:** Automated robotic systems and 'self-driving labs' enable high-throughput experimentation, parallel processing, and continuous operation, leading to substantial acceleration in areas like materials discovery and chemical synthesis (e.g., Burger 2020, Omidvar 2024 reported potential for 10-100x speedups; Arnold 2022 noted a 100x speedup in replicating a PhD project).

We analysed this emerging evidence and modeled potential acceleration under different AI capability levels. Based on documented achievements in specific, often optimized contexts, we derived plausible **upper-bound estimates for acceleration using future, highly advanced general-purpose AI systems ("Maximum-Level capabilities"):**

- **~100x acceleration** for primarily **Cognitive** research tasks.
- **~25x acceleration** for primarily **Physical** (experimental execution) tasks.

**The hypothetical future scenario for this expert elicitation:**

Please imagine the following — currently very futuristic — scenario:

1. **Maximum-level AI availability:** Very powerful general-purpose AI systems exist. They possess autonomous decision-making, advanced multi-disciplinary reasoning, and seamless integration with robotics. They can reliably plan, execute, analyze, interpret, and iterate on complex research cycles **with human-level capabilities and minimal human oversight, and at vastly increased speeds.**

2. **Universal access & use:** These powerful general-purpose AI systems are **accessible to everyone involved in the research process** – your group, collaborators, ethics committees, peer reviewers, journal editors, etc. They are striving towards delegating tasks where appropriate to achieve maximum effective acceleration while maintaining or improving quality.

**Your input:**

- Please consider the specific biomedical research project you have led.
- We will first ask about its actual timeline and task breakdown.
- Then, based on the hypothetical future scenario above, we want your expert judgment on the **realism** of achieving the **~100x (cognitive) / ~25x (physical)** maximum acceleration factors derived in our analysis, for each major research task within your project's context.
- We will also ask you to identify the key factors that you believe would limit acceleration, even with such powerful AI.

Your input will help better understand the true potential and practical boundaries of AI-driven research acceleration.
**Thank you for your participation!**

---

**Section A: Your Past Research Project Timeline**

Please think of **the specific biomedical research project** you substantially contributed to or led, from inception through to publication. Provide your best estimates for the questions below.

**1. Estimated total project duration (in months):** (From initial idea/synthesis to final publication)

[ ]

**2. Estimated time allocation (in months) across major research tasks:**
(Please ensure the months below add up to the total project duration entered above. These are estimates of primary focus time for your team) *

| [ ] | KNOWLEDGE SYNTHESIS (literature review, gap identification, etc) |
| [ ] | IDEA & HYPOTHESIS GENERATION (problem formulation, feasibility, etc) |
| [ ] | EXPERIMENT DESIGN (protocol development, methods selection, QC planning) |
| [ ] | ETHICS APPROVAL & PERMITS (documentation, review process, compliance) |
| [ ] | EXPERIMENT EXECUTION (lab work, sample prep, data acquisition, robotics) |
| [ ] | DATA ANALYSIS (cleaning, statistics, visualization, bioinformatics) |
| [ ] | RESULTS INTERPRETATION (synthesis of findings, hypothesis evaluation, context) |
| [ ] | MANUSCRIPT PREPARATION (writing, figures, references, revisions) |
| [ ] | PUBLICATION PROCESS (journal submission, peer review coordination, revisions) |

Total : 0

**Section B: Evaluating Maximum Potential Acceleration with Advanced AI**

Now, consider your project again within the **hypothetical future scenario** described in the Introduction (maximum-level AI, universal access and use). For each task, evaluate how plausible our estimated maximum acceleration factor seems i**f your specific project** had been carried out in the hypothetical future scenario.

**Task 1: Knowledge Synthesis**

*E.g., finding & curating information, critical evaluation, synthesizing findings, identifying gaps & contradictions*

Our analysis suggests a potential maximum acceleration of **~100x** for this cognitive task with Maximum-Level AI.

**How plausible is achieving this ~100x acceleration factor for this task in your project's context?**

- ○ Significant Overestimate (Plausible acceleration likely much lower)
- ○ Moderate Overestimate (Plausible acceleration likely moderately lower)
- ○ Plausible Estimate (Stated acceleration seems plausible)
- ○ Moderate Underestimate (Plausible acceleration likely moderately higher)
- ○ Significant Underestimate (Plausible acceleration likely much higher)

**Task 2: Idea & Hypothesis Generation**

*E.g., problem identification, hypothesis formulation, theoretical framework, feasibility assessment*

Our analysis suggests a potential maximum acceleration of **~100x** for this cognitive task with Maximum-Level AI

**How plausible is achieving this ~100x acceleration factor for this task in your project's context?**

- ○ Significant Overestimate (Plausible acceleration likely much lower)
- ○ Moderate Overestimate (Plausible acceleration likely moderately lower)
- ○ Plausible Estimate (Stated acceleration seems plausible)
- ○ Moderate Underestimate (Plausible acceleration likely moderately higher)

○ Significant Underestimate (Plausible acceleration likely much higher)

**Task 3: Experiment Design**

*E.g., method selection, protocol development, quality control planning*

Our analysis suggests a potential maximum acceleration of **~100x** for this cognitive task with Maximum-Level AI.

**How plausible is achieving this ~100x acceleration factor for this task in your project's context?**

○ Significant Overestimate (Plausible acceleration likely much lower)

○ Moderate Overestimate (Plausible acceleration likely moderately lower)

○ Plausible Estimate (Stated acceleration seems plausible)

○ Moderate Underestimate (Plausible acceleration likely moderately higher)

○ Significant Underestimate (Plausible acceleration likely much higher)

**Task 4: Ethics Approval & Permits**

*E.g., initial screening, scientific review, ethics assessment, regulatory compliance, administrative monitoring*

Our analysis suggests a potential maximum acceleration of **~100x** for this cognitive task with Maximum-Level AI, assuming institutional processes also adapt.

**How plausible is achieving this ~100x acceleration factor for this task in your project's context?**

○ Significant Overestimate (Plausible acceleration likely much lower)

○ Moderate Overestimate (Plausible acceleration likely moderately lower)

○ Plausible Estimate (Stated acceleration seems plausible)

○ Moderate Underestimate (Plausible acceleration likely moderately higher)

○ Significant Underestimate (Plausible acceleration likely much higher)

**Task 5: Experiment Execution**

*E.g., preparation, execution, documentation, troubleshooting & optimization, material management, equipment maintenance*

Our analysis suggests a potential maximum acceleration of **~25x** for this physical task with Maximum-Level AI and robotics/self-driving labs.

**How plausible is achieving this ~25x acceleration factor for this task in your project's context?**

○ Significant Overestimate (Plausible acceleration likely much lower)

○ Moderate Overestimate (Plausible acceleration likely moderately lower)

○ Plausible Estimate (Stated acceleration seems plausible)

○ Moderate Underestimate (Plausible acceleration likely moderately higher)

○ Significant Underestimate (Plausible acceleration likely much higher)

**Task 6: Data Analysis**

*E.g., data collection, cleaning & organization, statistical analysis, pattern identification, visualization, validation*

Our analysis suggests a potential maximum acceleration of **~100x** for this cognitive task with Maximum-Level AI.

**How plausible is achieving this ~100x acceleration factor for this task in your project's context?**

○ Significant Overestimate (Plausible acceleration likely much lower)

○ Moderate Overestimate (Plausible acceleration likely moderately lower)

○ Plausible Estimate (Stated acceleration seems plausible)

○ Moderate Underestimate (Plausible acceleration likely moderately higher)

○ Significant Underestimate (Plausible acceleration likely much higher)

**Task 7: Results Interpretation**

*E.g., synthesis of findings, hypothesis evaluation, contextualization within existing knowledge*

Our analysis suggests a potential maximum acceleration of **~100x** for this cognitive task with Maximum-Level AI.

**How plausible is achieving this ~100x acceleration factor for this task in your project's context?**

○ Significant Overestimate (Plausible acceleration likely much lower)

○ Moderate Overestimate (Plausible acceleration likely moderately lower)

○ Plausible Estimate (Stated acceleration seems plausible)

○ Moderate Underestimate (Plausible acceleration likely moderately higher)

○ Significant Underestimate (Plausible acceleration likely much higher)

---

**Task 8: Manuscript Preparation**

*E.g., methods documentation, results presentation, reference management, figure/table prep, data sharing, writing & revision*

Our analysis suggests a potential maximum acceleration of **~100x** for this cognitive task with Maximum-Level AI.

**How plausible is achieving this ~100x acceleration factor for this task in your project's context?**

○ Significant Overestimate (Plausible acceleration likely much lower)

○ Moderate Overestimate (Plausible acceleration likely moderately lower)

○ Plausible Estimate (Stated acceleration seems plausible)

○ Moderate Underestimate (Plausible acceleration likely moderately higher)

○ Significant Underestimate (Plausible acceleration likely much higher)

---

**Task 9: Publication Process**

*E.g., journal selection & submission, reviewer assignment, peer review, revision, correspondence*

Our analysis suggests a potential maximum acceleration of **~100x** for this (primarily) cognitive task with Maximum-Level AI, assuming institutional processes (journals, reviewers) also adapt.

**How plausible is achieving this ~100x acceleration factor for this task in your project's context?**

○ Significant Overestimate (Plausible acceleration likely much lower)

○ Moderate Overestimate (Plausible acceleration likely moderately lower)

○ Plausible Estimate (Stated acceleration seems plausible)

○ Moderate Underestimate (Plausible acceleration likely moderately higher)

○ Significant Underestimate (Plausible acceleration likely much higher)

---

**Section C: Evaluating Key Limiting Factors to Overall Research Acceleration**

In Section B, we asked you to evaluate the plausibility of potential maximum acceleration factors (~100x for cognitive tasks, ~25x for physical tasks) derived from optimistic scenarios in the literature. Often, practical limitations prevent achieving such theoretical maximums.

This section explores potential reasons why the actual acceleration achieved, even with Maximum-Level AI, might fall short of those ~100x/~25x figures.

For each factor below, please rate how significantly it contributes to **limiting the practical acceleration** achievable for the relevant research tasks (cognitive or physical), potentially explaining why the maximum factors presented earlier might be difficult to reach or sustain in a real-world project context.

Use the following scale:

- **Insignificant Limiter:** Unlikely to prevent achieving the maximum stated acceleration.
- **Minor Limiter:** May slightly reduce the practically achievable acceleration below the maximum.
- **Moderate Limiter:** Likely causes a noticeable reduction from the maximum achievable acceleration.
- **Major Limiter:** Likely a significant reason why the maximum acceleration won't be achieved.
- **Crucial Limiter:** Likely a primary reason why the maximum acceleration is unrealistic in practice.

**Please rate how significantly each factor limits the practical achievement of the maximum potential acceleration (~100x cognitive / ~25x physical) for the relevant tasks:**

---

### Fundamental Biological/Physical/Chemical Time & Measurement Limits

*E.g., irreducible time for biological processes, chemical reactions, physical equilibration; fundamental limits on measurement speed/sensitivity*

| Insignificant Limit | Minor Limit | Moderate Limit | Major Limit | Crucial Limit |
|---|---|---|---|---|
| ○ | ○ | ○ | ○ | ○ |

---

### Resource, Energy & Infrastructure Constraints

*E.g., availability, cost, supply chain, and total energy demands for materials, equipment, compute; time/cost for building/maintaining/upgrading labs/self-driving labs and compute infrastructure at scale*

| Insignificant Limit | Minor Limit | Moderate Limit | Major Limit | Crucial Limit |
|---|---|---|---|---|
| ○ | ○ | ○ | ○ | ○ |

---

### Input Data Limitations:

*E.g., constraints from quality, quantity, accessibility, biases, or time needed to generate novel ground truth data for AI.*

| Insignificant Limit | Minor Limit | Moderate Limit | Major Limit | Crucial Limit |
|---|---|---|---|---|
| ○ | ○ | ○ | ○ | ○ |

---

### Human Strategic Direction & Goal Setting

*E.g., need for human input to define research priorities, high-level goals, project scope, and criteria for "success"*

| Insignificant Limit | Minor Limit | Moderate Limit | Major Limit | Crucial Limit |
|---|---|---|---|---|
| ○ | ○ | ○ | ○ | ○ |

---

### Human Ethical Judgment & Value Alignment

*E.g., necessity for human oversight for complex ethical decisions, societal value alignment, and assessment of broader impacts*

| Insignificant Limit | Minor Limit | Moderate Limit | Major Limit | Crucial Limit |
|---|---|---|---|---|
| ○ | ○ | ○ | ○ | ○ |

---

### Human Accountability & Responsibility

*E.g., requirement for designated humans to hold ultimate responsibility for research conduct, outputs, and compliance*

| Insignificant Limit | Minor Limit | Moderate Limit | Major Limit | Crucial Limit |
|---|---|---|---|---|
| ○ | ○ | ○ | ○ | ○ |

---

### Institutional & Regulatory Adaptation:

*E.g., delays caused by the time needed for institutions (universities, journals, regulators) to adapt processes, standards, and legal frameworks*

| Insignificant Limit | Minor Limit | Moderate Limit | Major Limit | Crucial Limit |
|---|---|---|---|---|
| ○ | ○ | ○ | ○ | ○ |

---

### Empirical Validation & Inherent System Unpredictability

*E.g., time/resources for experimental testing of AI outputs; inherent unpredictability or complexity of the system under study necessitating iterative empirical cycles*

| Insignificant Limit | Minor Limit | Moderate Limit | Major Limit | Crucial Limit |
|---|---|---|---|---|
| ○ | ○ | ○ | ○ | ○ |

**Coordination & Consensus Among Human Stakeholders:**
*E.g., time delays from communication, deliberation, and agreement needed among essential human actors (collaborators, committees, editors, regulators)*

| Insignificant Limit | Minor Limit | Moderate Limit | Major Limit | Crucial Limit |
|---|---|---|---|---|
| ○ | ○ | ○ | ○ | ○ |

**Operational Safety, Security & Containment Assurance**
*E.g., time and procedures dedicated to ensuring the safe, secure, reliable, and contained operation of powerful autonomous research systems)*

| Insignificant Limit | Minor Limit | Moderate Limit | Major Limit | Crucial Limit |
|---|---|---|---|---|
| ○ | ○ | ○ | ○ | ○ |

**Scientific Community Assimilation & Conceptual Integration**
*E.g., limits on the rate the human scientific community can absorb, verify, conceptually integrate, and build upon a vastly increased volume of findings*

| Insignificant Limit | Minor Limit | Moderate Limit | Major Limit | Crucial Limit |
|---|---|---|---|---|
| ○ | ○ | ○ | ○ | ○ |

**Management of Extreme Data Volumes**
*E.g., potential bottlenecks in storage, transfer, curation, and accessibility arising from the sheer scale of data generated by hyper-accelerated research*

| Insignificant Limit | Minor Limit | Moderate Limit | Major Limit | Crucial Limit |
|---|---|---|---|---|
| ○ | ○ | ○ | ○ | ○ |

**Other Limiting Factors**
If you believe other significant limiting factors are missing, please specify below

[                    ]

**If you specified "Other" factors above, how significant a limit do you consider them, taken together?**
(Skip if you did not specify "Other" factors)

| Insignificant Limit | Minor Limit | Moderate Limit | Major Limit | Crucial Limit | N / A |
|---|---|---|---|---|---|
| ○ | ○ | ○ | ○ | ○ | ○ |

**Section D: General Considerations**

Please use this space for any additional thoughts:

- Any uncertainties or confusion you had while answering.
- Broader perspectives on how AI might transform research (beyond just speed).
- Concerns or potential risks associated with highly accelerated, AI-driven research.
- Ideas for overcoming the bottlenecks you identified to facilitate responsible acceleration.

[                    ]

**Section E: Contact Information**

**Name:** *

[                    ]

**Thank you for your time and valuable insights!**